\documentclass{iopart}

\usepackage{graphics}
\newcommand{\be}{\begin{equation}}
\newcommand{\ee}{\end{equation}}
\newcommand{\oln}[1]{\overline{#1}}
\newcommand{\uln}[1]{\underline{#1}}
\newcommand{\D}{\mathcal{D}}
\newcommand{\cH}{\mathcal{H}}
\newcommand{\R}{\mathbf{R}}

\newcommand{\diam}{\mathrm{diam}}
\newcommand{\Sch}{\mathrm{Sch}}
\newcommand{\ds}{\displaystyle}
\newcommand{\Flimit}[2]{{#1}\!\mbox{-}\!\lim_{#2}}
\newcommand{\dif}{\,\rmd}

\newtheorem{definition}{Definition}
\newtheorem{theorem}[definition]{Theorem}
\newtheorem{lemma}[definition]{Lemma}
\newtheorem{corollary}[definition]{Corollary}
\newcommand{\proof}{\par\noindent\emph{Proof: }}
\newcommand{\remark}{\par\noindent\emph{Remark: }}
\newcommand{\proofover}{\ensuremath{\bullet}\vspace{1em}}
\begin{document}
\paper{Calculus on fractal subsets of real line~--~I: formulation}
\author{Abhay Parvate and A D Gangal}
\address{Department of Physics, University of Pune}
\begin{abstract}
A new calculus based on fractal subsets of the
real line is formulated.
In this calculus, an integral of order
$\alpha, 0 < \alpha \leq 1$, called $F^\alpha$-integral, is defined, which
is suitable to integrate
functions with fractal support $F$ of dimension $\alpha$. Further, a derivative
of order $\alpha, 0 < \alpha \leq 1$, called $F^\alpha$-derivative, is
defined, which enables us to differentiate functions, like the Cantor
staircase, ``changing'' only on a fractal set. The $F^\alpha$-derivative is
local unlike the classical fractional derivative.
The $F^\alpha$-calculus retains much of the simplicity of ordinary calculus.
Several results including analogues of fundamental theorems of calculus are
proved.

The integral staircase function, which is a generalisation of the functions
like the Cantor staircase function, plays a key role in this formulation.
Further, it gives rise to a new definition of dimension, the
$\gamma$-dimension.

$F^\alpha$-differential equations are equations involving
$F^\alpha$-derivatives. They can be used to model sublinear dynamical
systems and fractal time processes, since sublinear behaviours are
associated with
staircase-like functions which occur naturally as their solutions.
As examples, we discuss a fractal-time diffusion equation, and one
dimensional motion of a particle undergoing friction in a fractal medium.

\end{abstract}

\section{Introduction}
It is now a well established fact that fractals can model many structures
found in nature~\cite{mandelbrot,Bunde}. The geometry of fractals is
also a well explored
subject~\cite{mandelbrot,falconer85,falconer90,falconer97,edgar}.

Fractals are often too irregular to have any smooth differentiable structure
defined on them, and render the methods and techniques of ordinary calculus
powerless or inapplicable. For example the derivative of a
Lebesgue-Cantor staircase function is zero almost everywhere and therefore
this function is not a solution of an ordinary differential equation.
Consequently, ordinary calculus does not equip us to handle problems such as
fractal time random walks, anomalous diffusion,
dynamics on fractals, fields of fractally
distributed sources etc., by setting up and solving ordinary differential
equations.

During recent times, a few approaches have been developed to deal with
various aspects of the problems mentioned above.

Several authors have recognized the need to use fractional derivatives and
integrals to explore the characteristic features of fractal walks, anomalous
diffusion, transport, etc.\ by setting up fractional kinetic equations,
master equations and so on~\cite{Metzler1994,Metzler1999,Hilfer1995,
Compte1996,zaslavsky1994,Metzler1999:2,Hilfer2000}. Fractional derivatives
are nonlocal operators and often are
suitable for modelling processes with memory but not always suitable to
handle the local scaling behaviour e.~g.\ the behaviour of fractal functions.
In~\cite{Kolwankar1996,Kolwankar1997,Kolwankar1998,Kolwankar1999}
this problem was circumvented by renormalising fractional derivatives and
constructing local fractional operators. This was further pursued
in~\cite{ Adda,Babakhani}. A particular success of this
approach was the demonstration of the striking fact that fractal and
multifractal functions
can be differentiated upto an order (fractional) determined by the Holder
exponent of the function (or dimension of its graph). In particular,
Weierstrass' nowhere differentiable function was
shown~\cite{Kolwankar1996,Kolwankar1997} to be differentiable
upto order $(1-\gamma)$, if $(1+\gamma)$ is the box dimension of its graph.

Another remarkable development is analysis on fractals. Many important ideas
and applications are developed in the realm of analysis on fractals. This
approach has been extensively used
for the treatment of diffusion, heat conduction, waves, etc.\ on
fractals---see \cite{Barlow,Kigami,Str:FDESG,Str:TASG} and several
references therein.

There is a further beautiful developement using a measure-theoretical
approach~\cite{FZ1,FZ2}. It
consists of defining derivative as the inverse of the integral with respect
to a measure and defining other operators using the derivative. This avoids
the dependence on the structure of the underlying fractal.

While all these themes have increased our understanding and brought out many
beautiful connections, a direct and simple approach involving fractional
order operators on fractal sets is only moderately explored.
Even though measure theoretical approach is elegant, Riemann integration
like procedures have their own place. They are more transparent,
constructive, and advantageous from algorithmic point of view.
It indeed seems possible to develope such an appropriate
calculus, tuned to these requirements. In the present paper, the first of
a series devoted to these ideas, we undertake a systematic developement of
calculus on fractal subsets of real line, involving integrals and
derivatives of appropriate orders $\alpha \in (0,1]$.
A brief glance at the table in~\ref{sec:analogies} would reveal that much
of the simplicity and intuitive appeal of ordinary calculus can be retained.

Differential equations of the form $D^\alpha f(x) = g(x)$, where $D^\alpha$
is a local differential operator of order $\alpha$, $0 < \alpha < 1$ were
considered in~\cite{Kolwankar1998}. It was argued that such equations can
have meaningful solutions if the dimension of the support of $g$ is
$\alpha$. Then the fractional integral of the characteristic function of
a Cantor set should be the corresponding Cantor staircase function, while
the local fractional derivative of the latter should be the former.
This was done using a Riemann integral like prescription. In particular a
fractional diffusion equation was shown to have subdiffusive solutions. This
enabled one to identify a new exact solution of Chapman-Kolmogorov equation.

This paper formulates the ideas of integral and derivative of
order $\alpha$, $0 < \alpha \leq 1$ based on a (fractal) set $F \subset \R$,
indicated in~\cite{Kolwankar1998}. We call them $F^\alpha$-integral and
$F^\alpha$-derivative respectively.

The organisation of the paper is as follows:
We begin in \sref{sec:staircase} by defining a mass function
$\gamma^\alpha(F,a,b)$ and integral staircase function $S^\alpha_F$
of an order $\alpha$ for a set $F$. The mass function
$\gamma^\alpha(F,a,b)$ gives us the content of a set~$F$ in the
interval~$[a,b] \subset \R$.
Its definition is based on Riemann-like sums. The construction can be
compared to the definition of Hausdorff measure, except that
the covers are more restrictive: they are in the form of finite
subdivisions of $[a,b]$. Though $\gamma^\alpha$ is not a
measure due to this simplification, it turns out to be proportional to
Hausdorff measure for compact sets.
The integral staircase function $S^\alpha_F(x)=\gamma^\alpha(F,a,x)$,
obtained from the mass
function by fixing $a$, is a generalization of the well known functions such
as the Lebesgue-Cantor staircase (or the Devil's staircase) functions.
The definitions of $F^\alpha$-integral and $F^\alpha$-derivative use
the quantity
$(S^\alpha_F(y)-S^\alpha_F(x))$ in place of the length $(y-x)$ of the interval
$[x,y]$.  In this respect, the definition of $F^\alpha$-integral is similar
to that of Riemann-Stieltjes integral~\cite{widder,shilov}.

In \sref{sec:gamma-dimension} it is shown that the mass function leads to
a new definition of dimension called $\gamma$-dimension, which is finer
than the box dimension, though not as fine as the Hausdorff dimension. In
later sections it is seen that the
$F^\alpha$-integral or the $F^\alpha$-derivative give meaningful results
when the $\gamma$-dimension of the underlying fractal $F$ is the same
as $\alpha$.

In \sref{sec:hausdorff} it is shown that the Hausdorff measure and the
mass function agree for compact sets upto a proportionality constant. Using
this property, the staircase function is calculated for the
middle~$\frac{1}{3}$ Cantor set.

Several sets can give rise to the same staircase function. A representative
set from such an equivalence class of sets, with nice properties, needs to be
chosen for defining the $F^\alpha$-derivative and proving the analogues of
fundamental theorems.
\Sref{sec:alpha-perfect} assures the existence and uniqueness of such a
set, called an $\alpha$-perfect set, associated with a staircase function.

We develope the rest of the theory in a way analogous to the standard
calculus~\cite{widder,shilov,goldberg}.
In \sref{sec:F-continuity} we introduce notations for limit and continuity
using the topology of $F$ with the metric inherited from $\R$. This
is done in order to distinguish them from limit and continuity on $\R$.

The ordinary integral of functions with fractal support $F \subset \R$ is
zero or undefined depending on the definition of integral (Lebesgue or
Riemann) and the nature of the support. The $F^\alpha$-integral defined in
\sref{sec:integration} suits the needs of integration of such functions.
It is further shown that the $F^\alpha$-integral of the characteristic
function of $F$ is the staircase function associated with $F$ as indicated
in~\cite{Kolwankar1998}.

Functions representing intermittent phenomena or fractal time evolution
typically ``change'' only on a fractal. The Cantor staircase function is an
example. The $F^\alpha$-derivative defined in \sref{sec:differentiation}
is best suited to quantify the ``rate of change'' of such functions.
This derivative is local unlike fractional
derivatives~\cite{Samko,Hilferbook2000,Miller,Oldham}. It is also not any
kind of average derivative as in~\cite{BF,PZ1,PZ2,PZ3,Z5}. It is more like
the first order derivative in ordinary calculus, which makes its dynamical
interpretation possible. Further in the same section, it is
shown that the $F^\alpha$-derivative of the staircase function of $F$ is the
characteristic function of $F$. Analogues of Rolle's theorem, the law of
the mean and Leibniz rule are discussed. In \sref{sec:fundamental-theorems},
we prove the analogues of fundamental theorems of calculus. The formula for
$F^\alpha$-integration by parts follows thereby.

The definitions of $F^\alpha$-integral and the $F^\alpha$-derivative reduce
to those of usual
Riemann integral and first order derivative respectively when $\alpha = 1$
and $F = \R$.

In \sref{sec:examples} we discuss examples including
subdiffusion and motion of a particle
undergoing friction in a fractal medium, and demonstrate the use of
$F^\alpha$-differential equations as their models.

As an example, the $F^\alpha$-integral of $f(x) = x\,\chi_C(x)$ for the
middle~$\frac{1}{3}$ Cantor set~$C$ is calculated in \ref{sec:x-chi}.
Repeated $F^\alpha$-derivatives and $F^\alpha$-integrals are discussed in
\ref{sec:repeated}, where we also calculate $F^\alpha$-derivatives and
$F^\alpha$-integrals of powers $(S^\alpha_F(x))^n$. A few analogies between
classical calculus and $F^\alpha$-calculus are tabulated in
\ref{sec:analogies}.

We begin by defining the integral staircase function.

\section{The mass function and the integral staircase}
\label{sec:staircase}

Let $F$ be a subset of the real line. In most of the interesting cases
discussed below, $F$ would be a fractal. In this section we
formulate the notion of the
content or $\alpha$-\emph{mass} of $F$ in an interval $[a,b]$, i.~e.\ mass of
$F \cap [a,b]$, of order $\alpha, 0<\alpha\leq 1$.

In all the following discussion, $0 < \alpha \leq 1$ unless stated
otherwise.

\begin{definition}
The \emph{flag function} $\theta(F,I)$ for a set $F$ and a closed
interval $I$ is given by
\be
\theta(F,I) = \cases{1 & if $F\cap I \neq \emptyset$\\
	0 & otherwise}
\label{eqn:theta}
\ee
\end{definition}

\begin{definition}
A \emph{subdivision} $P_{[a,b]}$ (or just $P$) of the interval
$[a,b]$, $a<b$, is a finite set of points
$\{a=x_0, x_1, \cdots, x_n=b\}$, $x_i < x_{i+1}$. Any interval of the
form $[x_i,x_{i+1}]$ is called a \emph{component interval} or just a
\emph{component} of the
subdivision~$P$. If $Q$ is any subdivision of $[a,b]$ and
$P \subset Q$, then we say that $Q$ is a \emph{refinement} of $P$.
If $a = b$, then the set $\{a\}$  is the only subdivision of $[a,b]$.
\end{definition}

\begin{definition}
For a set $F$ and a subdivision $P_{[a,b]}$, $a < b$,
\be
	\sigma^\alpha[F,P] = \sum_{i=0}^{n-1}
	\frac{(x_{i+1}-x_i)^\alpha}{\Gamma(\alpha+1)}
	\theta(F, [x_i,x_{i+1}]).
	\label{eqn:sigma}
\ee
If $a = b$, we define $\sigma^\alpha[F,P]$ to be zero.
\end{definition}
We note that the sum in \eref{eqn:sigma} contains a contribution from a
component interval if and only if that component contains at least one point
of $F$.  Further, $\sigma^\alpha[F,P] \geq 0$ for any set $F$
and subdivision $P$ of $[a,b]$.

We remark that this definition, and in particular
the factor $1/\Gamma(\alpha+1)$ and the use of finite subdivisions,
has been motivated by local fractional
calculus~\cite{Kolwankar1998,Kolwankar1999}.

Now we introduce the coarse-grained mass:
\begin{definition}
\label{def:gamma-delta}
Given $\delta > 0$ and $a \leq b$, the coarse-grained mass
$\gamma^\alpha_\delta(F,a,b)$ of $F \cap [a,b]$ is given by
\be
	\gamma^\alpha_\delta (F,a,b) =
	\inf_{\{P_{[a,b]}: |P| \leq \delta\}}
	\sigma^\alpha[F,P]
	\label{eqn:gamma-delta}
\ee
where
\be
	|P| = \max_{0 \leq i \leq n-1} (x_{i+1} - x_i)
	\label{eqn:largest-comp}
\ee
for a subdivision $P$,
and the infimum in \eref{eqn:gamma-delta} is taken over all
subdivisions $P$ of $[a,b]$ satisfying $|P| \leq \delta$.
\end{definition}

Eventually, a limit of $\gamma^\alpha_\delta(F,a,b)$ as
$\delta \rightarrow 0$ is taken in definition~\ref{def:gamma} below.
But before that
we examine some important properties of $\gamma^\alpha_\delta(F,a,b)$.

Let $a \leq b$ and $\delta_1 < \delta_2$. Then
$\gamma^\alpha_{\delta_1}(F,a,b)$ is the infimum of $\sigma^\alpha[F,P]$
over a smaller class of subdivisions than $\gamma^\alpha_{\delta_2}(F,a,b)$.
Thus:
\begin{lemma}
\label{lem:gamma-delta-delta}
Let $a \leq b$ and $\delta_1 < \delta_2$. Then
$\gamma^\alpha_{\delta_1}(F,a,b) \geq \gamma^\alpha_{\delta_2}(F,a,b)$.
\end{lemma}

The following lemma shows that
$\gamma^\alpha_\delta(F,a,b)$ is a monotonic increasing function of $b$
and a monotonic decreasing function of $a$. (Throughout the paper
we distinguish between \emph{monotonic} and \emph{strictly monotonic}.)
\begin{lemma}
\label{lem:gamma-delta-b}
Let $\delta > 0$ and $a < b < c$. Then, 
	$\gamma^\alpha_\delta(F,a,b) \leq \gamma^\alpha_\delta(F,a,c)$ and
	$\gamma^\alpha_\delta(F,b,c) \leq \gamma^\alpha_\delta(F,a,c)$.
\end{lemma}
\proof
Let $\epsilon > 0$. Then according to the definition~\ref{def:gamma-delta} of
$\gamma^\alpha_\delta(F,a,c)$, there exists a subdivision
$P_{[a,c]} = \{x_0=a,x_1,\dots,x_n=c\}$ such that $|P| \leq \delta$ and
\[
	\sigma^\alpha[F,P] < \gamma^\alpha_\delta(F,a,c) + \epsilon.
\]
Let $Q_{[a,b]} = \{x \in P: x < b\} \cup \{b\}$ i.~e.\
$Q_{[a,b]} = \{y_0, y_1, \dots, y_m\}$ where $y_i = x_i$ if $x_i < b$ and
$y_m = b$.

It follows that
$|Q_{[a,b]}| \leq |P_{[a,c]}| \leq \delta$ and
$\theta(F,[y_{m-1}, y_m]) \leq \theta(F,[x_{m-1},x_m])$ since
$[y_{m-1}, y_m] \subset [x_{m-1}, x_m]$. Therefore,
\[
	\sigma^\alpha[F,Q_{[a,b]}] \leq \sigma^\alpha[F,P_{[a,c]}]
	< \gamma^\alpha_\delta(F,a,c) + \epsilon.
\]
But $\gamma^\alpha_\delta(F,a,b) \leq \sigma^\alpha[F,Q]$ and
$\epsilon$ is arbitrary, hence
\[
	\gamma^\alpha_\delta(F,a,b) \leq \gamma^\alpha_\delta(F,a,c).
\]
which completes the proof of the first part. The second part follows in a
similar way. \proofover

\begin{theorem}
$\gamma^\alpha_\delta(F,a,b)$ is continuous in $b$ and $a$.
\label{thm:gamma-delta-continuity}
\end{theorem}
\proof
We prove the continuity of $\gamma^\alpha_\delta(F,a,b)$ in $b$ (with
$\delta$, $\alpha$ and $a$ fixed). Continuity
in $a$ can be proved in a similar manner.

Given $\epsilon>0$, let
\[
	\Delta' =
	\left({\epsilon\,\Gamma(\alpha+1)}\right)^\frac{1}{\alpha}
	\mbox{ and }
	\Delta = \min(\Delta',\delta).
\]

For $\epsilon_1>0$, there exists a subdivision $P$, 
such that $|P|\leq\delta$ and
\[
	\sigma^\alpha[F,P] < \gamma^\alpha_\delta(F,a,b) + \epsilon_1.
\]
Now
$Q = P \cup \{b+\Delta\}$ is a subdivision of $[a, b + \Delta]$. Therefore,
\begin{eqnarray*}
	\gamma^\alpha_\delta(F,a,b+\Delta)
	& \leq \sigma^\alpha[F,Q]\\
	& = \sigma^\alpha[F,P] +
	\theta(F, [b,b+\Delta])\frac{\Delta^\alpha}{\Gamma(\alpha+1)} \\
	& \leq \sigma^\alpha[F,P] + \epsilon\\
	& < \gamma^\alpha_\delta(F,a,b) + \epsilon_1 + \epsilon.
\end{eqnarray*}
As $\epsilon_1$ is arbitrary, we get $\gamma^\alpha_\delta(F,a,b+\Delta)
< \gamma^\alpha_\delta(F,a,b) + \epsilon$. As
$\gamma^\alpha_\delta(F,a,b)$ is a nondecreasing function of $b$,
\[
	\gamma^\alpha_\delta(F,a,b+t) < \gamma^\alpha_\delta(F,a,b) + \epsilon
\]
for $0 < t < \Delta$.

Summarizing, given $\epsilon>0$, there exists a $\Delta>0$
such that
\[
	c - b < \Delta \Longrightarrow
	\gamma^\alpha_\delta(F,a,c) - \gamma^\alpha_\delta(F,a,b) < \epsilon
\]
which implies that $\gamma^\alpha_\delta(F,a,b)$ is continuous in $b$ from
\emph{right}. The continuity from left follows on the replacement of $b$ by
$b-\Delta$ and of $b+\Delta$ by $b$ in the above proof.
\proofover

As mentioned earlier, the mass function is the limit of the coarse-grained
mass as $\delta \rightarrow 0$:
\begin{definition}
\label{def:gamma}
The mass function $\gamma^\alpha(F,a,b)$ is given by
\[
	\gamma^\alpha(F,a,b) =
	\lim_{\delta \rightarrow 0} \gamma^\alpha_\delta(F,a,b).
\]
\end{definition}
We note that since $\gamma^\alpha_\delta(F,a,b)$ increases as $\delta$
decreases,
$\gamma^\alpha(F,a,b)$ always exists and is a non-negative number, which may
possibly be~$+\infty$.

Another simple observation is that if $F \cap [a,b] = \emptyset$,
then $\gamma^\alpha_\delta(F,a,b)=0$ for any $\delta>0$, and consequently
$\gamma^\alpha(F,a,b)=0$. This result
can be extended so that it also applies to an open interval $(a,b)$:
\begin{lemma}
If $F\cap(a,b)=\emptyset$, then $\gamma^\alpha(F,a,b)=0$.
\end{lemma}
\proof
If $F \cap [a,b] = \emptyset$, then the result is obvious. If not, then
$F \cap [a,b]$ contains one or both of $a$ and $b$. In that case,
given $\epsilon>0$, we can choose a subdivision $P$ of $[a,b]$ such that
\[
	(x_1-x_0) \mbox{ and } (x_n-x_{n-1})
	< \left(\frac{\epsilon\,\Gamma(\alpha+1)}{2}\right)^\frac{1}{\alpha}
\]
where $\{x_0, \dots, x_n\}$ are points of $P$. Then,
$\sigma^\alpha[F,P] < \epsilon$ since $[x_1, x_{n-1}] \cap F = \emptyset$.
But as $\epsilon$ is arbitrary, it follows that
$\gamma^\alpha_\delta(F,a,b)=0$ for any $\delta>0$, so that
$\gamma^\alpha(F,a,b)=0$.
\proofover

A property, desired of a mass function, is additivity.
The following theorem asserts this.

\begin{theorem}
\label{thm:gamma-additivity}
Let $a < b < c$ and $\gamma^\alpha(F,a,c) < \infty$. Then
\be
	\gamma^\alpha(F,a,c) = \gamma^\alpha(F,a,b) + \gamma^\alpha(F,b,c).
	\label{eqn:gamma-additivity}
\ee
\end{theorem}
\proof
Given $\delta > 0$,
let $P_1$ be any subdivision of $[a,b]$ and $P_2$ be any subdivision of
$[b,c]$ such that $|P_1| \leq \delta$ and $|P_2| \leq \delta$. Then,
$P_1 \cup P_2$ is a subdivision of $[a,c]$, $|P_1 \cup P_2|\leq\delta$, and
\[
	\sigma^\alpha[F,P_1\cup P_2] = \sigma^\alpha[F,P_1] +
	\sigma^\alpha[F,P_2].
\]
Taking infimum over all subdivisions $P_1$ and $P_2$ such that
$|P_1|\leq\delta$
and $|P_2|\leq\delta$, and noting that \emph{not} all the subdivisions of
$[a,c]$ can be written in the form $P_1 \cup P_2$, where $P_1$ is a
subdivision of $[a, b]$ and $P_2$ is that of $[b,c]$, we get
\begin{eqnarray}
\gamma^\alpha_\delta(F,a,c)
& \leq
	\inf_{|P_1|\leq\delta,|P_2|\leq\delta} \sigma^\alpha[F,P_1\cup P_2]
	\nonumber \\
& =
	\gamma^\alpha_\delta(F,a,b) + \gamma^\alpha_\delta(F,b,c).
	\label{eqn:gamma-additivity-A}
\end{eqnarray}

Now for every subdivision $P_{[a,c]}$, $|P|\leq\delta$, we can construct a
subdivision $P' = P \cup \{b\}$. Obviously $|P'|\leq\delta$, and
$P' = P_1 \cup P_2$ where $P_1$
is a subdivision of $[a,b]$ and $P_2$ is a subdivision of $[b,c]$.

Let $P = \{x_0, x_1, \dots, x_n\}$.
If $b \in P$, then $P = P'$ and $\sigma^\alpha[F,P] = \sigma^\alpha[F,P']$.
Otherwise, let $[x_k,x_{k+1}]$ be the interval which contains $b$. Thus,
\begin{eqnarray*}
\fl	\sigma^\alpha[F,P\cup \{b\}] - \sigma^\alpha[F,P]
	& = & \theta(F,[x_k,b])
		\frac{(b - x_k)^\alpha}{\Gamma(\alpha+1)}
	\mbox{} + \theta(F,[b,x_{k+1}])
		\frac{(x_{k+1} - b)^\alpha}{\Gamma(\alpha+1)} \\
	& & \mbox{} - \theta(F,[x_k,x_{k+1}])
		\frac{(x_{k+1} - x_k)^\alpha}{\Gamma(\alpha+1)}.
\end{eqnarray*}
Hence,
\[
	\sigma^\alpha[F,P\cup \{b\}] - \sigma^\alpha[F,P] \leq
	\frac{3 \delta^\alpha}{\Gamma(\alpha+1)}.
\]
This implies that
\begin{eqnarray*}
	\sigma^\alpha[F,P] + \frac{3\delta^\alpha}{\Gamma(\alpha+1)}
& \geq &
	\sigma^\alpha[F,P\cup\{b\}] \\
& = &
	\sigma^\alpha[F,P_1] + \sigma^\alpha[F,P_2] \\
& \geq &
	\gamma^\alpha_\delta(F,a,b) + \gamma^\alpha_\delta(F,b,c)
\end{eqnarray*}
for all $P$. Thus if we take infimum over all subdivisions $P$ such that
$|P|\leq\delta$, we get
\be
	\gamma^\alpha_\delta(F,a,c) + \frac{3\delta^\alpha}{\Gamma(\alpha+1)}
	\geq
	\gamma^\alpha_\delta(F,a,b) + \gamma^\alpha_\delta(F,b,c)
	\label{eqn:gamma-additivity-B}
\ee
From \eref{eqn:gamma-additivity-A} and~\eref{eqn:gamma-additivity-B} and
taking the limit as $\delta\rightarrow 0$, we get the result.
\proofover

Since each term in \eref{eqn:gamma-additivity} is nonnegative for
$a \leq b \leq c$, an immediate consequence is
\begin{corollary}
$\gamma^\alpha(F,a,b)$ is increasing in $b$ and decreasing in $a$.
\label{cor:gamma-inc-dec}
\end{corollary}

The next theorem states that $\gamma^\alpha(F,a,x)$ takes all values in the
range $(0,\gamma^\alpha(F,a,b))$ for $x \in (a,b)$.

\begin{theorem}
Let $a<b$ and let $\gamma^\alpha(F,a,b)\neq 0$ be finite. Let $y$ be such
that $0 < y < \gamma^\alpha(F,a,b)$. Then there exists $c$, $a < c < b$, such
that $\gamma^\alpha(F,a,c) = y$.
\end{theorem}
\proof
Let $z = \gamma^\alpha(F,a,b) - y$.

Given a $\delta>0$, consider the set of all points $x$ of $[a,b]$ such that
$\gamma^\alpha_\delta(F,x,b) \leq z$. This set is an interval of the form
$[s_\delta,b]$ for some $s_\delta$, $a \leq s_\delta < b$, because
$\gamma^\alpha_\delta(F,x,b)$ is continuous
(theorem~\ref{thm:gamma-delta-continuity}) and decreasing
in $x$ (corollary~\ref{cor:gamma-inc-dec}). Since
$\gamma^\alpha_\delta(F,x,b)$
increases as $\delta$ decreases (lemma~\ref{lem:gamma-delta-delta}),
$s_\delta$ increases as $\delta$ decreases.

Similarly the set of all points $x$ of $[a,b]$ such that
$\gamma^\alpha_\delta(F,a,x) \leq y$ is an interval of the form
$[a,t_\delta]$, $a < t_\delta \leq b$, and $t_\delta$ decreases as
$\delta$ decreases.

Let $x \in (a,b)$. Then by theorem~\ref{thm:gamma-additivity},
\be
	\label{eqn:anyvalue-A}
	\gamma^\alpha(F,a,b)
	= \gamma^\alpha(F,a,x) + \gamma^\alpha(F,x,b)
	\geq
	\gamma^\alpha_\delta(F,a,x) +
	\gamma^\alpha_\delta(F,x,b).
\ee

As $y,z < \gamma^\alpha(F,a,b)$, there exists a $\delta_0 > 0$ such that
$\delta < \delta_0$ implies that $\gamma^\alpha_\delta(F,a,b) > y,z$. In the
rest of this proof, we only consider $\delta < \delta_0$
without mentioning.

Since $\gamma^\alpha_\delta(F,a,b) > y$ and $\gamma^\alpha_\delta(F,a,u)$ is
continuous and increasing in $u$, there exists an $x \in (a,b)$ such that
$\gamma^\alpha_\delta(F,a,x) = y$. This implies that $x \in [a, t_\delta]$.
Further, from \eref{eqn:anyvalue-A}, it follows that
\[
z = \gamma^\alpha(F,a,b) - y
= \gamma^\alpha(F,a,b) - \gamma_\delta^\alpha(F,a,x)
\geq \gamma^\alpha_\delta(F,x,b)
\]
implying that $x$ also belongs to $[s_\delta,b]$. This can happen only when
$s_\delta \leq t_\delta$.

Thus for each $\delta$ there exists an interval
$[s_\delta,t_\delta]$ such that
\[
	x \in [s_\delta,t_\delta] \Longrightarrow
	\gamma^\alpha_\delta(F,x,b) \leq z \mbox{ and }
	\gamma^\alpha_\delta(F,a,x) \leq y.
\]
Let $s = \sup_{0<\delta<\delta_0} s_\delta$ and
let $t = \inf_{0<\delta<\delta_0} t_\delta$.
Now $s_\delta$ increases and $t_\delta$ decreases as $\delta$ goes to zero,
but as $s_\delta \leq t_\delta$ for any $\delta$. Thus $s \leq t$ and
\[
	[s,t] = \bigcap_{0<\delta<\delta_0} [s_\delta,t_\delta].
\]
Consequently $x \in [s,t]$ implies $\gamma^\alpha_\delta(F,x,b) \leq z$
and $\gamma^\alpha_\delta(F,a,x) \leq y$ for any $\delta$. Hence
\be
	x \in [s,t] \Longrightarrow
	\gamma^\alpha(F,x,b) \leq z \mbox{ and }
	\gamma^\alpha(F,a,x) \leq y
	\label{eqn:anyvalue-B}
\ee
But as $ \gamma^\alpha(F,a,x) +
\gamma^\alpha(F,x,b) = \gamma^\alpha(F,a,b) = y + z$, the inequalities in
\eref{eqn:anyvalue-B} must be equalities. Thus for a given $y,
0<y<\gamma^\alpha(F,a,b)$, there exists a set $[s,t]\subset[a,b]$ such
that $x \in [s,t] \Longrightarrow \gamma^\alpha(F,a,x) = y$ which completes
the proof.
\proofover

\begin{corollary}
If $\gamma^\alpha(F,a,b)$ is finite, $\gamma^\alpha(F,a,x)$ is continuous
for $x \in (a,b)$.
\end{corollary}
This can be proved using the monotonicity of $\gamma^\alpha(F,a,b)$ in $a$
and $b$.
\remark
The implication of this result is that no single point has a nonzero mass,
or in other words, the mass function is atomless.

The scaling and translation properties of the mass function are similar to
those of Hausdorff measure:
\begin{theorem}
\label{thm:scaling}
For $F \subset \R$ and $\lambda \in \R$, let $F + \lambda$ denote the set
\[
	F + \lambda = \{x + \lambda: x \in F\}
\]
and let $\lambda F$ denote the set
\[
	\lambda F = \{\lambda x: x \in F\}.
\]
Then,
\begin{enumerate}
\item Translation:
\[
	\gamma^\alpha(F+\lambda, a+\lambda, b+\lambda)
	= \gamma^\alpha(F,a,b)
\]
\item Scaling ($\lambda \geq 0$):
\[
	\gamma^\alpha(\lambda F, \lambda a, \lambda b)
	= \lambda^\alpha\,\gamma^\alpha(F,a,b)
\]
\end{enumerate}
\end{theorem}
\remark
If the set $F$ is self-similar so that
$\lambda_0 F \cap [\lambda_0 a, \lambda_0 b]
= F \cap [\lambda_0 a, \lambda_0 b]$ for a particular $\lambda_0$,
then the scaling property can be rewritten as
\[
	\gamma^\alpha(F,\lambda_0 a, \lambda_0 b)
	= \lambda_0^\alpha \gamma^\alpha(F,a,b).
\]
An example is the middle~$\frac{1}{3}$ Cantor set $C$, with $a=0,b=1$ and
$\lambda_0 = \frac{1}{3^n}$.

Now we introduce one of the central notions of this paper,
viz.\ the integral staircase function for a
set~$F$ of the order $\alpha$. This function, which is a generalization of
functions like the Lebesgue-Cantor staircase function,
describes how the mass of $F\cap[a,b]$ increases as $b$ increases.
\begin{definition}
\label{def:staircase}
Let $a_0$ be an arbitrary but fixed real number.
The integral staircase function $S^\alpha_F(x)$ of order $\alpha$ for a
set $F$ is given by
\be
	S^\alpha_F(x) = \cases{
		\gamma^\alpha(F,a_0,x) & if $x \geq a_0$ \\
		-\gamma^\alpha(F,x,a_0) & otherwise.
	}
	\label{eqn:staircase}
\ee
\end{definition}
The number $a_0$ can be chosen according to convenience.
A few properties of $S^\alpha_F(x)$ which are restatements of the
corresponding properties of the mass function $\gamma^\alpha(F,a,b)$ are as
follows.
\begin{theorem}
\label{thm:restatements}
Let $F$ be a subset of $\R$, and let $0 < \alpha \leq 1$. If
$\gamma^\alpha(F,a,b)$ is
finite, then for all $x, y \in (a,b)$ such that $x < y$, the following
statements hold:
\begin{enumerate}
\item
	$S^\alpha_F(x)$ is increasing in $x$.
\item
	If $F\cap(x,y) = \emptyset$, then $S^\alpha_F$ is a
	constant in $[x,y]$.
\item
	$S^\alpha_F(y) - S^\alpha_F(x) = \gamma^\alpha(F,x,y)$.
\item
	$S^\alpha_F$ is continuous on $(a,b)$.
\end{enumerate}
\end{theorem}

As an example, we calculate and show the graph of $S^\alpha_C$ for the
middle $\frac{1}{3}$ Cantor set $C$ in the \sref{sec:hausdorff}, after
discussing some results required to calculate it.

\section{The $\gamma$-dimension}
\label{sec:gamma-dimension}

We now consider the sets $F$ for which the mass function
$\gamma^\alpha(F,a,b)$ gives the
most useful information.
Due to the similarity of the definitions of the mass function and the
Hausdorff outer
measure~\cite{mandelbrot,falconer85,falconer90,falconer97,edgar},
one might expect that
the mass function can be used to define a fractal dimension. It is indeed
the case. If $0<\alpha<\beta\leq 1$,
\[
	\sigma^\beta[F,P] \leq
	|P|^{\beta-\alpha} \sigma^\alpha[F,P]
	\frac{\Gamma(\alpha+1)}{\Gamma(\beta+1)}
\]
so that
\[
	\gamma_\delta^\beta(F,a,b) \leq
	\delta^{\beta-\alpha} \gamma_\delta^\alpha(F,a,b)
	\frac{\Gamma(\alpha+1)}{\Gamma(\beta+1)}
\]
Thus in the limit as $\delta \rightarrow 0$, we get
\[
	\gamma^\beta(F,a,b) = 0
	\qquad \mbox{provided } \gamma^\alpha(F,a,b) < \infty \mbox{ and }
	\alpha < \beta.
\]
It follows that $\gamma^\alpha(F,a,b)$ is infinite upto certain value of
$\alpha$, say $\alpha_0$, and jumps down to zero $\alpha > \alpha_0$
(if $\alpha_0 < 1$).
We call this number the $\gamma$-dimension of $F$.
$\gamma^{\alpha_0}(F,a,b)$ itself may be zero, nonzero finite,
or infinite. To make the notion of dimension precise,
\begin{definition}
The $\gamma$-dimension of $F\cap[a,b]$, denoted by
$\dim_\gamma(F\cap[a,b])$, is
\begin{eqnarray*}
	\dim_\gamma(F\cap[a,b]) & = &
		\inf \{\alpha:\gamma^\alpha(F,a,b) = 0\} \\
	& = &
		\sup \{\alpha:\gamma^\alpha(F,a,b) = \infty\}
\end{eqnarray*}
\end{definition}

Now we compare the $\gamma$-dimension with the Hausdorff dimension and the
box dimension. As the definition of Hausdorff measure
involves arbitrary countable covers,
it is expected that the Hausdorff dimension be finer than
$\gamma$-dimension. This is shown to be the case below:

Let $\cH^\alpha_\delta(E)$ denote the coarse grained Hausdorff measure of
a subset $E$ of $\R$, and $\cH^\alpha(E)$ denote the Hausdorff measure.
Let $P$ be a subdivision with $|P|\leq\delta$.
Those components $[x_i,x_{i+1}]$ of $P$ for which $\theta(F,[x_i,x_{i+1}])$
is nonzero, form a $\delta$-cover of $F\cap[a,b]$. Thus,
\[
	\sigma^\alpha[F,P]
	\geq \frac{1}{\Gamma(\alpha+1)} \cH^\alpha_\delta(F\cap[a,b]).
\]
Since this is true for any $P$ such that $|P| \leq \delta$, it follows that
\[
	\gamma_\delta^\alpha(F,a,b) \geq
	\frac{1}{\Gamma(\alpha+1)} \cH_\delta^\alpha(F\cap[a,b])
\]
for each $\delta>0$. So taking limit as $\delta \rightarrow 0$,
\be
	\gamma^\alpha(F,a,b) \geq
	\frac{1}{\Gamma(\alpha+1)} \cH^\alpha(F\cap[a,b]).
	\label{eqn:hausdorff-gamma}
\ee
which also implies
\[
	\dim_{\cH}(F\cap[a,b]) \leq
	\dim_\gamma(F\cap[a,b]).
\]

There exist sets for which the two definitions give different results. For
example, if $\mathbf{Q}$ denotes the set of rational numbers, then
$\dim_\cH(\mathbf{Q}\cap[0,1])=0$, while
$\dim_\gamma(\mathbf{Q}\cap[0,1])=1$. However, it will be shown in the next
section that the two dimensions are equal for compact sets.

Next we compare the $\gamma$-dimension with the box dimension.
Let $\dim_\gamma(F\cap[a,b]) = \alpha$. Then $\gamma^\beta(F,a,b)$ diverges
for any $\beta < \alpha$. Thus for any $k > 0$, there exists $\delta_0>0$
such that
$\delta < \delta_0 \Longrightarrow \gamma^\beta_\delta(F,a,b) > k$.

Let $P$ be any subdivision such that $|P|\leq\delta$, and let
$N_\delta(F\cap[a,b])$ be the number of nonzero terms in the sum
$\sigma^\alpha[F,P]$. Then, for arbitrary but fixed $k>0$ and
$\delta < \delta_0$,
\[
	k < \gamma^\beta_\delta(F,a,b) \leq
	\frac{N_\delta(F\cap[a,b]) \delta^\beta}{\Gamma(\beta+1)}
\]
where $0 < \beta < \alpha \leq 1$. Thus,
\[
	\ln(k) \leq \ln N_\delta(F\cap[a,b]) + \beta \ln(\delta)
	- \ln(\Gamma(\beta+1))
\]
which implies
\[
	-\beta \ln(\delta) \leq \ln N_\delta(F\cap[a,b]) - \ln(k) -
	\ln(\Gamma(\beta+1)).
\]
Dividing by $-\ln(\delta)$ (which is positive for $\delta<1$),
\[
	\beta \leq
	\frac{\ln(N_\delta(F\cap[a,b]))}{-\ln(\delta)}
	- \frac{\ln(k) - \ln(\Gamma(\beta+1))}{-\ln(\delta)}
\]
Taking limit as $\delta \rightarrow 0$ and noting that the first term is the
definition of the box dimension $\dim_B(F\cap[a,b])$ in the limit and the
denominator of the second diverges, we get
\[
	\beta \leq
	\dim_B(F\cap[a,b]) =
	\lim_{\delta \rightarrow 0}
	\frac{\ln(N_\delta(F\cap[a,b]))}{-\ln(\delta)}.
\]
This is true for any $\beta<\alpha = \dim_\gamma(F\cap[a,b])$, so that
\[
	\dim_\gamma(F\cap[a,b]) \leq \dim_B(F\cap[a,b]).
\]

As an example in which box dimension and $\gamma$-dimension differ, we
consider the set $K=\{0,1,\frac{1}{2},\frac{1}{3},\dots\}$.
It is known~\cite{falconer90} that $\dim_B K = 0.5$, and $\dim_\cH K = 0$.
Further, as $K$ is compact, $\dim_\gamma K = \dim_\cH K$ according to the
corollary~\ref{cor:compact-dimension} to be proved in the next section.

Thus the $\gamma$-dimension is finer than the box dimension, but not than
Hausdorff dimension. Specifically, the $\gamma$-dimension is unaffected by
clusters of points unlike the box dimension, as the above example of the set
$K$ demonstrates. On the other hand, it is sensitive to countable but dense
sets such as rationals.

The modified box dimension~\cite{falconer90} involves countable subsets of
the set under
consideration, to circumvent the above mentioned problems in box dimension.
Therefore the former is finer than $\gamma$-dimension. But the
$\gamma$-dimension is simpler to calculate.
For similar reasons, the $\gamma$-dimension is simpler than packing
dimension~\cite{falconer90}.

Summarizing, if $E \subset \R$, then
\[
	\dim_B E \geq \dim_\gamma E \geq \dim_{MB} E \geq \dim_\cH E.
\]

\section{The mass function and the Hausdorff measure}
\label{sec:hausdorff}

For compact sets, any open covers, and in particular countable open covers,
can be replaced by finite subcovers. So the mass function and Hausdorff measure are expected to be proportional.
Theorem~\ref{thm:compact-equality} below shows that this is indeed the case.
In what follows, $\cH^\alpha_\delta(F)$ denotes the coarse grained Hausdorff
measure~\cite{falconer85} of a set $F \subset \R$, and $\cH^\alpha(F)$
denotes the Hausdorff measure of $F$, both of order~$\alpha$.

\begin{theorem}
\label{thm:compact-equality}
For a compact set $F \subset \R$,
\[
	\gamma^\alpha(F,a,b) =
	\frac{1}{\Gamma(\alpha+1)} \cH^\alpha(F\cap[a,b]).
\]
\end{theorem}
\proof
For $\delta > 0$, let $\{A_i, i = 1,2,\dots\}$ be any countable cover of
$F\cap[a,b]$ such that $\diam A_i \leq \frac{\delta}{2}$ for all $i$. The
sets $A_i$ need not be open or closed. Then
\[
	\cH^\alpha_{\delta/2}(F\cap[a,b]) \leq \sum_i (\diam A_i)^\alpha.
\]
Consider closed intervals $B_i = [u_i, v_i]$ where $u_i = \inf A_i$ and
$v_i = \sup A_i$. Then $A_i \subset B_i$ and $\diam B_i = \diam A_i$. Thus
$\{B_i\}$ forms a cover of $F\cap[a,b]$ and
\[
	\sum_i (\diam B_i)^\alpha = \sum_i (\diam A_i)^\alpha
	\geq \cH^\alpha_{\delta/2}(F\cap[a,b]).
\]
Given $\epsilon \in (0, (\delta/2)^\alpha)$, let $\{C_i\}_{i=1}^\infty$ be
the open intervals
\[
	C_i = \left(
	u_i - \frac{1}{2}
	\left(\frac{\epsilon}{2^{i}}\right)^\frac{1}{\alpha},\ 
	v_i + \frac{1}{2}
	\left(\frac{\epsilon}{2^{i}}\right)^\frac{1}{\alpha}
	\right).
\]
The class $\{C_i\}_{i=1}^\infty$ thus forms an open cover of $F\cap[a,b]$
and
\[
	\diam C_i = \diam A_i
	+ \left( \frac{\epsilon}{2^{i}} \right)^\frac{1}{\alpha}
	< \delta
\]
so that
\[
	\sum (\diam C_i)^\alpha
	 = \sum \left( \diam A_i +
	\left(\frac{\epsilon}{2^{i}}\right)^\frac{1}{\alpha}
	\right)^\alpha .
\]
A simple consequence of Jensen's inequality~\cite{beckenbach}, which for the
case of two variables assures that $(s_1+s_2)^t \leq s_1^t + s_2^t$ for
$s_1,s_2>0$ and $0<t<1$, is that
\be
	\sum (\diam C_i)^\alpha
	\leq \sum (\diam A_i)^\alpha + \epsilon \sum\frac{1}{2^{i}}
	= \sum (\diam A_i)^\alpha + \epsilon.
	\label{eqn:compact-one}
\ee

We now show that a finite cover consisting of closed intervals can be
constructed.
As $F$ is compact, so is $F\cap[a,b]$. Thus a finite subset of $\{C_i\}$
covers $F\cap[a,b]$. We denote this finite subcover by
$\{D_i, i = 1, \dots, n\}$. The $D_i$ are open intervals of the form $(a_i,
b_i)$. Without loss of generality we can choose this finite subcover
$\{D_i\}$ such that $D_i \not\subset D_j$ whenever $i \neq j$. Further, the
sets are labeled such that $a_i \leq a_{i+1}$. But as
$D_i \not\subset D_{i+1}$ and $D_{i+1} \not\subset D_i$, it implies that
$a_i < a_{i+1}$ and $b_i < b_{i+1}$.

Now we consider the closures $\oln{D}_i$ of $D_i$. As $\{D_i\}$ is a finite
subcover out of $\{C_i\}$ and $\{\oln{D}_i\}$ have the same diameters as
$D_i$, it follows from \eref{eqn:compact-one} that
\[
	\sum (\diam \oln{D}_i)^\alpha \leq \sum (\diam A_i)^\alpha +
	\epsilon.
\]
Let $I_1 = \oln{D}_1$ and $I_i = \oln{D}_i \backslash D_{i-1}$
for $2 \leq i \leq n$.
The collection $\{I_i\}$ forms a finite cover of $F\cap[a,b]$ by closed
intervals, and
\[
	\sum (\diam I_i)^\alpha \leq \sum (\diam A_i)^\alpha + \epsilon.
\]
The closed intervals $I_i$ share at the most endpoints. The set of
all the endpoints of $I_i$, $1 \leq i \leq n$ forms a subdivision
$P$ of $[a,b]$ which can be refined to a subdivision $Q$ such that
$|Q| \leq \delta$ and
\[
	\Gamma(\alpha+1) \sigma^\alpha[F,Q] = \sum (\diam I_i)^\alpha
	\leq \sum (\diam A_i)^\alpha + \epsilon.
\]
Therefore,
\[
	\Gamma(\alpha+1) \gamma^\alpha_\delta (F,a,b)
	\leq \sum (\diam A_i)^\alpha + \epsilon.
\]
Since this relation holds for any countable cover $\{A_i\}$ of
$F\cup[a,b]$ such that $\diam A_i \leq \delta/2$ and for arbitrary
$\epsilon > 0$, it follows that
\[
	\Gamma(\alpha+1) \gamma^\alpha_\delta(F,a,b) \leq
	\cH^\alpha_{\delta/2}(F\cap[a,b]) \leq
	\cH^\alpha(F\cap[a,b]).
\]
Consequently in the limit as $\delta \rightarrow 0$,
\be
	\label{eqn:compact-two}
	\Gamma(\alpha+1) \gamma^\alpha(F,a,b) \leq
	\cH^\alpha(F\cap[a,b]).
\ee
Equations~(\ref{eqn:hausdorff-gamma}) and~(\ref{eqn:compact-two})
together imply the required equality.
\proofover
\begin{corollary}
\label{cor:compact-dimension}
If $F \subset \R$ is compact, then $\dim_\gamma F = \dim_\cH F$.
\end{corollary}

\begin{figure}[htb]
\begin{center}%
\resizebox{0.6\columnwidth}{!}%
	{\includegraphics{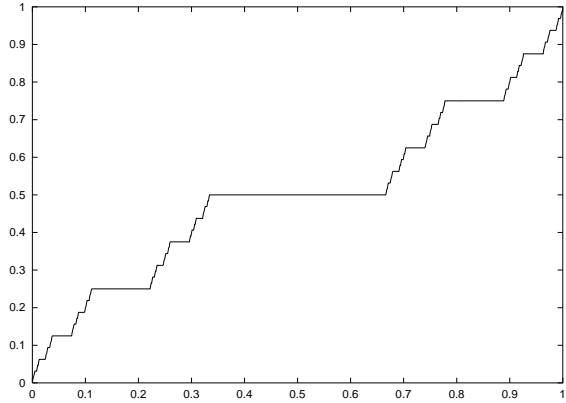}}%
\end{center}
\caption{\label{fig:cantor-staircase}%
	$\Gamma(\alpha+1) S^\alpha_C(x)$:
	The integral staircase function for the Cantor set}
\end{figure}

\emph{Example}: We now discuss an important prototype example of
$S^\alpha_F(x)$.  Consider the middle 1/3 Cantor set $C$ (hereafter
referred to as the Cantor set).
This set is compact and has a Hausdorff dimension $\alpha=\log(2)/\log(3)$.
Thus by theorem~\ref{thm:compact-equality},
$\cH^\alpha(C\cap[a,b]) = \Gamma(\alpha+1)\gamma^\alpha(C,a,b)$ and
$\dim_\gamma C = \dim_\cH C = \alpha = \log(2)/\log(3)$.
Using the self-similarity of $C$ and the monotonicity as well as scaling and
translation properties of the mass function (theorem~\ref{thm:scaling}),
we can calculate $S^\alpha_F$ at each point.  A graph of
$\Gamma(\alpha+1) S^\alpha_C(x)$ i.~e.\
$\Gamma(\alpha+1) \gamma^\alpha(C,0,x)$
against $x$ is shown in \fref{fig:cantor-staircase}. This is the
Lebesgue-Cantor Staircase function.

\section{$\alpha$-Perfect sets}
\label{sec:alpha-perfect}

The correspondence between sets $F$ and their staircase functions
$S^\alpha_F$ is many to one. For example, $S^\alpha_{C'} = S^\alpha_C$ where
$C$ is the middle $\frac{1}{3}$ Cantor set and
$C' = C\backslash\{\frac{2}{3}\}$. Intuitively, it can be
said that as the mass function is atomless, removing a single point from $C$
does not change its value.

Another example is a set
$D = C \cup E$ where $E \subset (\frac{1}{3}, \frac{2}{3})$ satisfying
$\dim_\gamma E < \alpha$, $\alpha = \ln(2)/\ln(3)$.
Then it can be seen that $S^\alpha_D(x) = S^\alpha_C(x)$ for all $x$. Thus
adding a lower dimensional set need not change the value of the staircase
function either.  We call the sets giving rise to the same staircase function
as staircasewise congruent:

\begin{definition}
Let $F \subset \R$ and $G \subset \R$ be such that
$\dim_\gamma F = \dim_\gamma G =\alpha$, $\alpha \in (0,1]$. Then $F$ and
$G$ are said to be staircasewise congruent if $S^\alpha_F(x)$ and
$S^\alpha_G(x)$ are finite and equal for all $x \in \R$.

This congruence being an equivalence relation, we denote the equivalence
class of sets
containing $F$ by $\mathcal{E}^\alpha_F$.  Thus if $G$ is in
$\mathcal{E}^\alpha_F$, then $S^\alpha_G = S^\alpha_F$ and
$\mathcal{E}^\alpha_G = \mathcal{E}^\alpha_F$.
\end{definition}

The above examples intuitively suggest that not all points or subsets
contribute to the staircase function. Now we proceed to
select a representative set out of the equivalence class which, intuitively
speaking, has exactly those points at which $S^\alpha_F$ ``changes''.
To choose only the points where a function ``changes'', we need the
following definition.
\begin{definition}
We say that a point $x$ is a point of change of a function $f$, if $f$ is
not constant over any open interval $(c,d)$ containing $x$. The set of all
points of change of $f$ is called the set of change of $f$ and is denoted by
$\Sch f$.
\end{definition}

Thus $\Sch f_1 = \emptyset$ if $f_1 = \mbox{constant}$, while
$\Sch f_2 = \R$ if $f_2(x) = x$. More importantly we note that if 
$G \in \mathcal{E}^\alpha_F$, then $S^\alpha_G(x) = S^\alpha_F(x)$ and
therefore $\Sch(S^\alpha_G) = \Sch(S^\alpha_F)$. In other words,
$\Sch(S^\alpha_F)$ is determined by the equivalence class
$\mathcal{E}^\alpha_F$.

The following theorem states that $\Sch(S^\alpha_F)$ itself belongs to
$\mathcal{E}^\alpha_F$.

\begin{theorem}
\label{thm:sch-belongs}
Let $F \subset \R$ be such that $S^\alpha_F(x)$ is finite for all $x \in \R$
for $\alpha = \dim_\gamma F$ and $H = \Sch(S^\alpha_F)$. Then $H$ belongs to
$\mathcal{E}^\alpha_F$ i.~e. $S^\alpha_H = S^\alpha_F$.
\end{theorem}
\proof
It will be sufficient, in view of definition~\ref{def:staircase}, to prove
that $\gamma^\alpha(H,a,b) = \gamma^\alpha(F,a,b)$ for any $a,b \in \R$.

We begin by noting that for $u < v$, if $F\cap[u,v] = \emptyset$, then
$\gamma^\alpha(F,u,v) = 0$.
Consequently $S^\alpha_F$ is constant on $(u,v)$ implying $(u,v)\cap H =
\emptyset$.
Then for any $\epsilon' > 0$ such that $u + \epsilon' \leq v-\epsilon'$,
\be
	\theta(H, [u+\epsilon', v-\epsilon']) = 0.
	\label{eqn:sch-belongs-A}
\ee

Next, let $\delta>0$. Then given $\epsilon > 0$, there is a subdivision
$P_{[a,b]} = \{y_0, y_1, \dots, y_n\}$ such that $|P| \leq \delta$ and
\be
	\label{eqn:alpha-perfect-aa}
	\sigma^\alpha[F,P]
	\leq \gamma^\alpha_\delta(F,a,b) + \frac{\epsilon}{2}.
\ee
If $\theta(F,I) = 1$ for all
components $I$ of $P$, then certainly
\be
	\label{eqn:alpha-perfect-a}
	\sigma^\alpha[H,P] \leq \sigma^\alpha[F,P].
\ee
Otherwise, let $K$ be the set of all points of the form
\be
	\label{eqn:alpha-perfect-ab}
	c' = c +
	\left(\frac{\epsilon\,\Gamma(\alpha+1)}{2n}\right)^\frac{1}{\alpha},\
	d' = d -
	\left(\frac{\epsilon\,\Gamma(\alpha+1)}{2n}\right)^\frac{1}{\alpha}
\ee
where $c$ and $d$ are the endpoints of those components $I$ of $P$ such that
$\theta(F,I) = 0$. Then $Q_{[a,b]} = P \cup K$ is a refined subdivision
and $|Q| \leq |P|$. If $I=[c,d]$ is a component of $P$ such that
$\theta(F,I)=0$, then it contains three components of $Q$, viz.\ $[c,c']$,
$[c',d']$ and $[d',d]$ where $c'$ and $d'$ are given by
\eref{eqn:alpha-perfect-ab}. The term in $\sigma^\alpha[H,Q]$
corresponding to $[c',d']$ is zero according to
\eref{eqn:sch-belongs-A}, and the remaining two contribute at the
most $\epsilon/2n$ each. If $\theta(F,I) \neq 0$, then $I$ is also a
component of $Q$ and the term corresponding to $I$ in $\sigma^\alpha[H,Q]$
is either zero or is exactly the same as the corresponding term
in $\sigma^\alpha[F,P]$. Therefore,
\be
	\label{eqn:alpha-perfect-b}
	\sigma^\alpha[H,Q] \leq \sigma^\alpha[F,P] + \frac{\epsilon}{2}.
\ee
Thus from \eref{eqn:alpha-perfect-aa}, \eref{eqn:alpha-perfect-a}
and~\eref{eqn:alpha-perfect-b},
we see that there exists a subdivision $Q$, such that $|Q| \leq \delta$ and
\[
	\sigma^\alpha[H,Q] \leq \gamma^\alpha_\delta(F,a,b) + \epsilon.
\]
As $\epsilon$ is arbitrary, we see that
\be
	\label{eqn:alpha-perfect-d}
	\gamma^\alpha_\delta(H,a,b) \leq \gamma^\alpha_\delta(F,a,b)
\ee

Now we wish to rule out the possibility that
\be
	\label{eqn:alpha-perfect-da}
	\gamma^\alpha_\delta(H,a,b) < \gamma^\alpha_\delta(F,a,b).
\ee
Suppose that \eref{eqn:alpha-perfect-da} is true. Then there exists a
subdivision $P_1 = \{x_0, \dots, x_n\}$ such that $|P_1| \leq \delta$ and
\be
	\label{eqn:alpha-perfect-e}
	\sigma^\alpha[H,P_1] < \gamma^\alpha_\delta(F,a,b).
\ee
From \eref{eqn:gamma-additivity-A} we know that
\[
	\gamma^\alpha_\delta(F,a,b) \leq
	\sum_{i=0}^{n-1} \gamma^\alpha_\delta(F, x_i, x_{i+1})
\]
so that
\[
	\sigma^\alpha[H,P_1] <
	\sum_{i=0}^{n-1} \gamma^\alpha_\delta(F, x_i, x_{i+1}).
\]
For this equation to hold, there must be at least one
$k$, $0 \leq k \leq n-1$, such that
\[
	\frac{(x_{k+1} - x_k)^\alpha}{\Gamma(\alpha+1)}
	\theta(H,[x_k, x_{k+1}]) <
	\gamma^\alpha_\delta(F,x_k, x_{k+1}) \leq
	\gamma^\alpha(F,x_k, x_{k+1}).
\]
As every quantity is non-negative, it follows that
$\gamma^\alpha(F, x_k, x_{k+1}) > 0$, and $S^\alpha_F$ is not constant in
$[x_k, x_{k+1}]$. Therefore
$\Sch(S^\alpha_F)\cap[x_k,x_{k+1}] \neq \emptyset$ i.~e.\
$H\cap[x_k,x_{k+1}]\neq\emptyset$. So $\theta(H,[x_k,x_{k+1}])=1$ and
\[
	\frac{(x_{k+1} - x_k)^\alpha}{\Gamma(\alpha+1)} <
	\gamma^\alpha_\delta(F,x_k, x_{k+1}).
\]
As $Q = \{x_k, x_{k+1}\}$ is a subdivision of $[x_k, x_{k+1}]$ such
that $|Q| \leq \delta$,
\[
	\frac{(x_{k+1} - x_k)^\alpha}{\Gamma(\alpha+1)}
	= \sigma^\alpha[F,Q] <
	\gamma^\alpha_\delta(F,x_k, x_{k+1})
\]
which is a contradiction by the definition of
$\gamma^\alpha_\delta(F,x_k, x_{k+1})$ implying that our assumption
\eref{eqn:alpha-perfect-da} is wrong. Thus,
\eref{eqn:alpha-perfect-d} is an equality
for any $\delta > 0$, and therefore
$\gamma^\alpha(H,a,b) = \gamma^\alpha(F,a,b)$.
\proofover

\begin{lemma}
Let $F \subset \R$ be such that $S^\alpha_F(x)$ is finite for all $x\in\R$
for $\alpha = \dim_\gamma F$. Then the set $H = \Sch(S^\alpha_F)$ is perfect
i.~e.\ $H$ is closed and every point of $H$ is its limit point.
\end{lemma}
\proof
Let $y$ be a limit point of $H$. Then any open interval $(c,d)$ containing
$y$ contains a point $z$ of $H = \Sch(S^\alpha_F)$. Therefore $S^\alpha_F$
is not constant on $(c,d)$. Hence $y \in H$ implying that $F$ is closed.

If $x \in H$ is not a limit point of $H$, then there exists an open interval
$(c,d)$ containing $x$ but no other point of $H$ so that
$F \cap (c,x) = \emptyset$ and $F\cap(x,d) = \emptyset$. This implies that
$S^\alpha_F(d) - S^\alpha_F(c) =
(S^\alpha_F(d) - S^\alpha_F(x)) + 
(S^\alpha_F(x) - S^\alpha_F(c)) = 0$ due to
theorem~\ref{thm:restatements}(ii).
Therefore $x$ is not in $H = \Sch(S^\alpha_F)$ which is a contradiction.
\proofover

Now we choose $\Sch(S^\alpha_F)$ as the ``canonical'' representative of
$\mathcal{E}^\alpha_F$:
\begin{definition}
Let $F \subset \R$ be such that $S^\alpha_F(x)$ is finite for all $x\in\R$
for $\alpha = \dim_\gamma F$. Then the set $\Sch(S^\alpha_F)$ is said to be
$\alpha$-perfect, and is said to be the $\alpha$-perfect representative of
$\mathcal{E}^\alpha_F$.
\end{definition}
Thus, every $\mathcal{E}^\alpha_F$ contains a unique $\alpha$-perfect set.
The next theorem states that it is the minimal closed set in
$\mathcal{E}^\alpha_F$.

\begin{theorem}
An $\alpha$-perfect set $F$ is the intersection of all the
closed sets $G$ in $\mathcal{E}^\alpha_F$. In other words, it is the minimal
closed set in $\mathcal{E}^\alpha_F$.
\end{theorem}
\proof
Let $\mathcal{G}$ be the class of all closed sets in $\mathcal{E}^\alpha_F$.

As $F$ is perfect, it is closed. Therefore,
\be
	\label{eqn:minimal-a}
	F \supset \bigcap_{G \in \mathcal{G}} G.
\ee

Let $G_0 \in \mathcal{G}$ and $x \notin G_0$. Then there is an open interval
$(c,d)$ containing $x$ but no point of $G_0$, as $G_0$ is closed. This
implies that
$S^\alpha_{G_0}(c) = S^\alpha_{G_0}(d)$ by
theorem~\ref{thm:restatements}(ii), and
further $S^\alpha_F(c) = S^\alpha_F(d)$ as $G_0 \in \mathcal{E}^\alpha_F$.
Since $F$ is $\alpha$-perfect, we have $F\cap(c,d)=\emptyset$ implying that
$x \notin F$. Therefore, $x \in F \Longrightarrow x \in G_0$ for all
$G_0 \in \mathcal{G}$, so that
\be
	\label{eqn:minimal-b}
	F \subset \bigcap_{G \in \mathcal{G}} G.
\ee
The proof is completed in view of \eref{eqn:minimal-a}
and \eref{eqn:minimal-b}.
\proofover

The following lemma is a useful restatement of the definition of an
$\alpha$-perfect set.
\begin{lemma}
\label{lem:oneside}
Let $F \subset \R$ be $\alpha$-perfect and $x\in F$. If
$y<x<z$, then either $S^\alpha_F(y) < S^\alpha_F(x)$ or
$S^\alpha_F(x) < S^\alpha_F(z)$ (or both).
\end{lemma}
Thus for an $\alpha$-perfect set $F$, the lemma assures that if
$x \in F$, then the values of $S^\alpha_F(y)$ must be different from
$S^\alpha_F(x)$ at all points~$y$ on at least one side of $x$.

As an example, we now show that the middle $\frac{1}{3}$ Cantor set $C$ is
$\alpha$-perfect, for $\alpha = \log(2)/\log(3)$. Let $x \notin C$. As $C$
is closed, there is at least one open interval $(c,d)$ containing $x$, such
that $C\cap(c,d)=\emptyset$. Therefore, $S^\alpha_C$ is constant on $(c,d)$
implying that $x \notin \Sch(S^\alpha_C)$. Hence,
\be
	\label{eqn:alpha-perfect-exa}
	\Sch(S^\alpha_C) \subset C.
\ee

Let $x$ be a point of $C$. Then $x$ can be represented by $0.x_1x_2x_3\dots$
where $x_i$ is the $i$th digit in the ternary representation of $x$.
As $x \in C$, $x_i = 0 \mbox{ or } 2$. Let $(c,d)$ be any open interval
containing $x$. Then there is an integer $n>0$ such that
$(x - 3^{-n}, x + 3^{-n}) \subset (c,d)$.
Let $D$ be the set of numbers $y = 0.y_1y_2y_3\dots$ satisfying
\[
	y_i=\cases{
		x_i & $1 \leq i \leq n+1$\\
		0 \mbox{ or } 2 & $i > n+1$
	}
\]
where $y_i$ is the $i$th digit in the ternary representation of $y$.
Then $D\subset C$ is a scaled down
copy of $C$ by a factor $3^{-n-1}$ and $D \subset (x - 3^{-n}, x + 3^{-n})
\subset (c,d)$. Thus using the scaling property of $\gamma^\alpha$
(theorem~\ref{thm:scaling}),
\[
	S^\alpha_C(d) - S^\alpha_C(c)
	\geq \frac{1}{\Gamma(\alpha+1)} (3^{-n-1})^\alpha
	= \frac{1}{\Gamma(\alpha+1)} 2^{-n-1} > 0
\]
where $\alpha = \ln(2)/\ln(3)$.
This implies that $S^\alpha_C$ is not constant on $(c,d)$. Thus, $x \in C
\Longrightarrow x \in \Sch(S^\alpha_C)$ i.~e.
\be
	\label{eqn:alpha-perfect-exb}
	C \subset \Sch(S^\alpha_C).
\ee
From \eref{eqn:alpha-perfect-exa} and~\eref{eqn:alpha-perfect-exb} we see
that $C = \Sch(S^\alpha_C)$ implying that $C$ is $\alpha$-perfect for
$\alpha = \log(2)/\log(3)$.

\section{$F$-continuity}
\label{sec:F-continuity}

In this section we introduce the notation for limit and continuity using
topology of $F \subset \R$ with the metric inherited from $\R$.
Our purpose in doing so is to distinguish between these notions and ones
on~$\R$ when they both appear together.

\begin{definition}
Let $F \subset \R$, $f: \R \rightarrow \R$ and $x \in F$. A number $\ell$ is
said to be the limit of $f$ through the points of $F$, or simply
$F$-limit, as $y \rightarrow x$, if given any $\epsilon > 0$, there exists
$\delta > 0$ such that
\[
	y \in F \mbox{ and }|y-x| < \delta
	\Longrightarrow
	|f(y) - \ell| < \epsilon.
\]
If such a number exists, then it is denoted by
\[
	\ell = \Flimit{F}{y \rightarrow x} f(y)
\]
\end{definition}
This definition does not involve values of the function at $y$ if
$y \notin F$. Also, $F$-limit is not defined at points $x \notin F$.

We now introduce the notion of $F$-continuity which is continuity as
far as the values of the function only on the set $F$ are concerned.

\begin{definition}
A function $f:\R \rightarrow \R$ is said to be $F$-continuous at $x \in F$ if
\[
	f(x) = \Flimit{F}{y \rightarrow x} f(y)
\]
\end{definition}
We note that the notion of $F$-continuity is not defined at $x \notin F$.

It is clear that continuity of $f: \R \rightarrow \R$ at $x \in F$ implies
$F$-continuity at $x$. But the converse is not
true. We consider a few examples: Let $C$ be the middle $\frac{1}{3}$ Cantor
set. Then the functions $f_1(x) = 1$ and $f_2(x) = x$ are continuous on
$[0,1]$; they are also $C$-continuous on $C\cap[0,1]$. In contrast, consider
$f_3(x) = \chi_C(x)$ and $f_4(x) = x \cdot \chi_C(x)$ where $\chi_C(x)$ is
the characteristic function of $C$. These functions are $C$-continuous, but
not continuous.

Now we define an analogue of uniform continuity:

\begin{definition}
A function $f:\R \rightarrow \R$ is said to be uniformly
$F$-continuous on $E \subset F$ if for any $\epsilon>0$ there
exists $\delta>0$ such that
\[
	x \in F, y \in E \mbox{ and } |y - x|<\delta \Longrightarrow
	|f(y)-f(x)|<\epsilon.
\]
\end{definition}
It is clear that uniform $F$-continuity on $E$ implies $F$-continuity on
$E$. The converse is true only in certain cases:

\begin{theorem}
If a function $f:\R \rightarrow \R$ is $F$-continuous on a compact set
$E \subset F$, then it is uniformly $F$-continuous on $E$.
\end{theorem}

\section{$F^\alpha$-Integration}
\label{sec:integration}

In the definition of $F^\alpha$-integral below, values of the function
only at the points of $F$
are considered. Further, instead of the lengths of subintervals, we consider
the difference between the values of the staircase function $S^\alpha_F$ at
the endpoints. In this respect, $F^\alpha$-integral is similar to
Riemann-Stiltjes integral~\cite{widder,shilov}.

\begin{definition}
The class of functions $f: \R \rightarrow \R$ which are bounded on $F$
is denoted by $B(F)$. In other words,
\[
	f \in B(F) \Longleftrightarrow
	-\infty < \inf_{x \in F} f(x)
	\leq \sup_{x \in F} f(x) < +\infty
\]
\end{definition}

As the first step, we now define upper and lower sums which approximate the
value of the $F^\alpha$-integral.

\begin{definition}
\label{def:Mm}
Let $f \in B(F)$. Let $I$
be a closed interval. Then,
\begin{eqnarray*}
	M[f,F,I] & = \sup_{x \in F \cap I} f(x)
		\qquad & \mbox{if } F \cap I \neq \emptyset \\
		& = 0 \qquad & \mbox{otherwise}
\end{eqnarray*}
and similarly
\begin{eqnarray*}
	m[f,F,I] & =  \inf_{x \in F \cap I} f(x)
		\qquad & \mbox{if } F \cap I \neq \emptyset \\
		& = 0 \qquad & \mbox{otherwise}
\end{eqnarray*}
\end{definition}

\begin{definition}
Let $S^\alpha_F(x)$ be finite for $x \in [a,b]$.
Let $P$ be a subdivision of $[a,b]$ with points $x_0,\dots,x_n$.
The upper $F^\alpha$-sum and the lower $F^\alpha$-sum for the function $f$
over the subdivision $P$ are given respectively by
\be
	U^\alpha[f,F,P] = \sum_{i=0}^{n-1} M[f,F,[x_i,x_{i+1}]]
	(S^\alpha_F(x_{i+1}) - S^\alpha_F(x_i))
	\label{eqn:u-alpha}
\ee
and
\be
	L^\alpha[f,F,P] = \sum_{i=0}^{n-1} m[f,F,[x_i,x_{i+1}]]
	(S^\alpha_F(x_{i+1}) - S^\alpha_F(x_i))
	\label{eqn:l-alpha}
\ee
\end{definition}
We emphasize the appearance of intersection $F \cap I$ in the definition of
$M$ and $m$, and also the use of $(S^\alpha_F(x_{i+1}) - S^\alpha_F(x_i))$
as in a Riemann-Stieltjes sum instead of $(x_{i+1} - x_i)$.

From the definition it is clear that
\be
	U^\alpha[f,F,P] \geq L^\alpha[f,F,P].
	\label{eqn:U-geq-L}
\ee

The following lemma asserts that with refinements, the upper $F^\alpha$-sum
decreases and the lower $F^\alpha$-sum increases, both monotonically.
\begin{lemma}
\label{lem:subdiv-refine}
Let $F\subset\R$ and $f \in B(F)$.
If $Q$ is a refinement of a subdivision $P$, then
$U^\alpha[f,F,Q] \leq U^\alpha[f,F,P]$ and
$L^\alpha[f,F,Q] \geq L^\alpha[f,F,P]$.
\end{lemma}
\proof
To start with, let $P = \{x_0, x_1, \dots, x_n\}$ and
$Q = P \cup \{x'\}$ where $x' \in (x_i, x_{i+1})$.
Let $I=[x_i,x_{i+1}]$,
$I'=[x_i,x']$, and $I''=[x',x_{i+1}]$. If
there are no points of $F$ in $I$,
then $M[f,F,I]=M[f,F,I']=M[f,F,I'']=0$. Otherwise there
are two possibilities; either both $I'$ and $I''$ have
points of $F$, or only one of them, say $I'$ without loss of
generality, has points of $F$.

In the first case,
\[
	M[f,F,I'] \leq M[f,F,I] \mbox{ and }
	M[f,F,I''] \leq M[f,F,I].
\]
Thus we have,
\begin{eqnarray}
	\fl M[f,F,I] (S^\alpha_F(x_{i+1}) - S^\alpha_F(x_i)) \nonumber\\
\lo =
	M[f,F,I]
	\{(S^\alpha_F(x') - S^\alpha_F(x_i)) + (S^\alpha_F(x_{i+1}) -
	S^\alpha_F(x'))\} \nonumber \\
\lo \geq
	M[f,F,I'] (S^\alpha_F(x') - S^\alpha_F(x_i)) +
	M[f,F,I''] (S^\alpha_F(x_{i+1}) - S^\alpha_F(x')).
	\label{eqn:UL-refine-1}
\end{eqnarray}

In the second case, only $I'$ has the points of $F$. Consequently
$S^\alpha_F(x_{i+1}) = S^\alpha_F(x')$ and $M[f,F,I'] = M[f,F,I].$
Thus
\be
	M[f,F,I] (S^\alpha_F(x_{i+1}) - S^\alpha_F(x_i))
	= M[f,F,I'] (S^\alpha_F(x') - S^\alpha_F(x_i))
	\label{eqn:UL-refine-2}
\ee
since $M[f,F,I''] = 0$.
Combining \eref{eqn:UL-refine-1} and~\eref{eqn:UL-refine-2}, we have
\begin{eqnarray*}
	\fl M[f,F,I] (S^\alpha_F(x_{i+1}) - S^\alpha_F(x_i)) \\
	\lo\geq
	M[f,F,I'] (S^\alpha_F(x') - S^\alpha_F(x_i)) +
	M[f,F,I''] (S^\alpha_F(x_{i+1}) - S^\alpha_F(x')).
\end{eqnarray*}
Thus
\[
	U^\alpha[f,F,Q] \leq U^\alpha[f,F,P].
\]
This conclusion can easily be extended for any refinement of $P$.

By a similar argument, we can prove that
\[
	L^\alpha[f,F,Q] \geq L^\alpha[f,F,P]
\]
which completes the proof.
\proofover

\begin{lemma}
If $P$ and $Q$ are any two subdivisions of $[a,b]$, then
\[
	U^\alpha[f,F,P] \geq L^\alpha[f,F,Q]
\]
\end{lemma}
\proof
As $P \cup Q$ is a refinement of both $P$ and $Q$, it follows from the above
lemma and \eref{eqn:U-geq-L} that
\[
	U^\alpha[f,F,P] \geq U^\alpha[f,F,P \cup Q] \geq L^\alpha[f,F,P \cup
	Q] \geq L^\alpha[f,F,Q].
\]
\proofover

We are now ready to define the $F^\alpha$-integral.
\begin{definition}
Let $F$ be such that $S^\alpha_F$ is finite on $[a,b]$.
For $f \in B(F)$, the lower $F^\alpha$-integral is given by
\be
	\uln{\int_a^b}f(x)\dif^\alpha_F x = \sup_{P_{[a,b]}} L^\alpha[f,F,P]
	\label{eqn:lower-int}
\ee
and the upper $F^\alpha$-integral is given by
\be
	\oln{\int_a^b}f(x)\dif^\alpha_F x = \inf_{P_{[a,b]}} U^\alpha[f,F,P]
	\label{eqn:upper-int}
\ee
Both the supremum and infimum are taken over all the subdivisions $P$ of
$[a,b]$.
\end{definition}
The $\dif^\alpha_F x$ appearing in \eref{eqn:lower-int}
and~\eref{eqn:upper-int} has no separate meaning; it is just the
notation.

It is obvious that
\be
	\uln{\int_a^b}f(x)\dif^\alpha_F x \leq
	\oln{\int_a^b}f(x)\dif^\alpha_F x
	\label{eqn:lower-upper-int}
\ee

\begin{definition}
If $f \in B(F)$, we say that $f$ is $F^\alpha$-integrable on $[a,b]$ if
\[
	\uln{\int_a^b}f(x)\dif^\alpha_F x =
	\oln{\int_a^b}f(x)\dif^\alpha_F x
\]
In that case the $F^\alpha$-integral of $f$ on $[a,b]$, denoted by
$\int_a^b f(x)\dif^\alpha_F x$ is given by the common value.
\end{definition}

For future use we note the following obvious and useful criterion for
proving $F^\alpha$-integrability:
\begin{lemma}
\label{lem:epsilon-int}
Let $f \in B(F)$. Then $f$ is $F^\alpha$-integrable on $[a,b]$ if
and only if, for any $\epsilon>0$,
there exists a subdivision $P$ of $[a,b]$ such that
\[
	U^\alpha[f,F,P] < L^\alpha[f,F,P] + \epsilon.
\]
\end{lemma}

Now we state a sufficient condition for $F^\alpha$-integrability. The
sufficient and necessary conditions will be discussed in a companion paper.
\begin{theorem}
Let $F$ be such that $F\cap[a,b]$ is compact and $S^\alpha_F$ is finite on
$[a,b]$. Let $f \in B(F)$, and $a<b$. If $f$ is
$F$-continuous on $F\cap[a,b]$, then $f$ is $F^\alpha$-integrable on
$[a,b]$.
\end{theorem}
\proof
If $S^\alpha_F(a) = S^\alpha_F(b)$, then the $F^\alpha$-integral is zero,
and the result is obvious.

Now consider the case where $S^\alpha_F(a) \neq S^\alpha_F(b)$.
The function $f$ is uniformly $F$-continuous on $F\cap[a,b]$ as $F\cap[a,b]$
is compact.  Thus, given $\epsilon>0$, there is a $\delta>0$ such that
\be
	\fl
	\label{eqn:cont-int}
	x,y \in F\cap[a,b] \mbox{ and } |y-x| < \delta
	\Longrightarrow
	|f(y)-f(x)| < \frac{\epsilon}{S^\alpha_F(b)-S^\alpha_F(a)}.
\ee
Let $P$ be a subdivision such that $|P| < \delta$. Then it can be seen that
$U^\alpha[f,F,P] < L^\alpha[f,F,P] + \epsilon$ which completes the proof in
view of lemma~\ref{lem:epsilon-int}.
\proofover

The following property of $F^\alpha$-integral is expected from any fair
definition of an integral:
\begin{theorem}
Let $a<b$ and $f$ be an $F^\alpha$-integrable function on $[a,b]$.
Let $c \in (a,b)$.
Then, $f$ is $F^\alpha$-integrable on $[a,c]$ and $[c,b]$.
Further,
\be
	\int_a^b f(x)\dif^\alpha_F x =
	\int_a^c f(x)\dif^\alpha_F x +
	\int_c^b f(x)\dif^\alpha_F x
	\label{eqn:int_ac_cb}
\ee
\end{theorem}
This can be proved in a manner analogous to Riemann integral.

The linearity of $F^\alpha$-integral follows from the definition:
\begin{theorem}
\begin{enumerate}
\item
If $f$ is $F^\alpha$-integrable on $[a,b]$, and $\lambda$ is
any real number, then
\[
	\int_a^b \lambda f(x)\dif^\alpha_F x =
	\lambda \int_a^b f(x)\dif^\alpha_F x.
\]
\item
If $f$ and $g$ are $F^\alpha$-integrable functions
on $[a,b]$, then
\[
	\int_a^b (f(x)+g(x))\dif^\alpha_F x =
	\int_a^b f(x)\dif^\alpha_F x +
	\int_a^b g(x)\dif^\alpha_F x.
\]
\end{enumerate}
\end{theorem}

The following lemma states an obvious property:
\begin{lemma}
If $f$ and $g$ are $F^\alpha$-integrable over $[a,b]$, and
$f(x)\geq g(x)$ for all $x \in F \cap [a,b]$, then
\[
	\int_a^b f(x)\dif^\alpha_F x \geq
	\int_a^b g(x)\dif^\alpha_F x
\]
\end{lemma}

\begin{definition}
If $f$ is $F^\alpha$-integrable on $[a,b]$, $a<b$, then
\[
	\int_b^a f(x)\dif^\alpha_F x =
	- \int_a^b f(x)\dif^\alpha_F x.
\]
\end{definition}

A particularly simple but important example, as realized
in~\cite{Kolwankar1998}, is the $F^\alpha$-integral of the characteristic
function $\chi_F$ of the set $F$:
\begin{lemma}
\label{lem:chi-integral}
If $\chi_F(x)$ is the characteristic function of $F \subset \R$, then
\[
	\int_a^b \chi_F(x)\dif^\alpha_F x
	= S^\alpha_F(b) - S^\alpha_F(a)
\]
\end{lemma}
\proof
For a closed interval $I \subset [a,b]$,
\begin{eqnarray*}
	M[\chi_F,F,I] = m[\chi_F,F,I]
	& = 1 \qquad & \mbox{if } F \cap I \neq \emptyset \\
	& = 0 \qquad & \mbox{otherwise}
\end{eqnarray*}
so that $M[\chi_F,F,I]$ is zero for a closed interval $I = [c,d]$ only when
$S^\alpha_F(d)-S^\alpha_F(c)=0$. Thus
\[
	U^\alpha[\chi_F,F,P] = L^\alpha[\chi_F,F,P] = S^\alpha_F(b) - S^\alpha_F(a)
\]
for any subdivision $P$ of $[a,b]$.
\proofover

As a further example, in \ref{sec:x-chi} we calculate the
$C^\alpha$-integral of the function $f(x) = x\,\chi_C(x)$ where $C$ is the
middle~$\frac{1}{3}$ Cantor set, and $\alpha = \ln(2)/\ln(3)$ is its
$\gamma$-dimension. The integral is given by equations~\eref{eqn:x-pre-nine}
and~\eref{eqn:x-nine}.

\section{$F^\alpha$-Differentiation}
\label{sec:differentiation}

Like the first
order derivative, the $F^\alpha$-derivative is a limit of a quotient. But
here the limit is $F$-limit, and the denominator is the difference in
the values of the staircase function $S_F^\alpha$ at two points. Moreover,
intuitively speaking, $F$ is typically the set of change of the function,
and $\alpha$ is typically the $\gamma$-dimension of $F$.

\begin{definition}
\label{def:derivative}
If $F$ is an $\alpha$-perfect set
then the $F^\alpha$-derivative of $f$ at $x$ is
\be
	\D^\alpha_F (f(x)) =
	\cases{
		\Flimit{F}{y \rightarrow x}
		\frac{f(y)-f(x)}{S^\alpha_F(y)-S^\alpha_F(x)}
	& if $x \in F$\\
	0 & otherwise}
	\label{eqn:def-derivative}
\ee
if the limit exists.
\end{definition}
Note that lemma~\ref{lem:oneside} tells us that if $x \in F$, then we would
find such points $y$ which are arbitrarily close to $x$ at least on one
side of $x$ so that the denominator in the definition is not zero and the
RHS in \eref{eqn:def-derivative} makes sense.

We now state a necessary condition for the above limit to exist.

\begin{theorem}
If $\D^\alpha_F(f(x))$ exists for all $x$ in $(a,b)$, then $f(x)$ is
$F$-continuous in $(a,b)$.
\end{theorem}
The proof is straightforward.

The linearity of the $F^\alpha$-derivative is an immediate consequence of
the definition~\ref{def:derivative}. Thus:

\begin{theorem}
\label{thm:derivative-linearity}
\begin{enumerate}
\item
Let $f$ be a function on $[a,b]$. If $\D^\alpha_F(f(x))$ exists for all
$x \in [a,b]$, then $\D^\alpha_F(\lambda f(x))$ exists and
\[
	\D^\alpha_F(\lambda f(x)) = \lambda \D^\alpha_F(f(x)).
\]
\item
Let $f$ and $g$ be functions on $[a,b]$. If $\D^\alpha_F(f(x))$ and
$\D^\alpha_F(g(x))$ exist for all $x \in [a,b]$, then
$\D^\alpha_F(f(x)+g(x))$ exists and
\[
	\D^\alpha_F(f(x)+g(x)) = \D^\alpha_F(f(x)) + \D^\alpha_F(g(x)).
\]
\end{enumerate}
\end{theorem}

Now we calculate the derivative for two elementary functions.
The first does not need a proof:
\begin{lemma}
The $F^\alpha$-derivative of a constant function $f: \R \rightarrow \R$,
$f(x) = k \in \R$ is zero, i.~e.
\[
	\D^\alpha_F(f) = 0.
\]
\end{lemma}
This result is to be contrasted with the classical fractional derivative
(Riemann-Liouville, and others) of a constant which is not zero in
general~\cite{Samko,Hilferbook2000,Miller,Oldham}.

\begin{lemma}
\label{lem:staircase-derivative}
The derivative of the integral staircase itself is the characteristic
function $\chi_F$ of $F$:
\[
	\D^\alpha_F(S^\alpha_F(x)) = \chi_F(x).
\]
\end{lemma}
\proof
If $x \notin F$, $\D^\alpha_F(S^\alpha_F(x)) = 0$.

If $x \in F$, then
\[
	\D^\alpha_F(S^\alpha_F(x)) = 
	\Flimit{F}{y \rightarrow x}
	\frac{S^\alpha_F(y)-S^\alpha_F(x)}{S^\alpha_F(y)-S^\alpha_F(x)}
	= 1
\]
\proofover

This lemma together with lemma~\ref{lem:chi-integral} can be viewed as the
special cases of the fundamental theorems of calculus
(\sref{sec:fundamental-theorems}) involving $S^\alpha_F$
and its derivative $\chi_F$.

An analogue of Rolle's theorem is:
\begin{theorem}
Let $f: \R \rightarrow \R$ be a continuous function such that
$\Sch f \subset F$ where $F$ is $\alpha$-perfect,
$\D^\alpha_F(f(x))$ is defined for all $x \in [a,b]$, and
$f(a) = f(b) = 0$. Then there is a point $c \in F\cap[a,b]$ such that
$\D^\alpha_F(f(c)) \geq 0$ and a point $d \in F\cap[a,b]$ such that
$\D^\alpha_F(f(d)) \leq 0$.
\end{theorem}
\proof
If $f$ is zero throughout $[a,b]$, then $\D^\alpha_F(f(x))=0$ for all
$x \in [a,b]$ and the result follows in this case.

If $f(y)>0$ for some $y \in (a,b)$, then as $f$ is continuous, there exists
an open interval $(c,d)\subset(a,b)$ containing $y$ such that
$f(z) > 0$ for any $z \in (c,d)$. Let $(c_0,d_0)$ be largest such
interval. Then $f(c_0) = f(d_0) = 0$.
The point $c_0 \in \Sch f \subset F$ as $f$ is positive on the right of
$c_0$.
So,
\[
	\D^\alpha_F(c_0) =
	\Flimit{F}{z \rightarrow c_0}
	\frac{f(z)-f(c_0)}{S^\alpha_F(z)-S^\alpha_F(c_0)}
	\geq 0.
\]
Similarly $d_0 \in F$ and
\[
	\D^\alpha_F(d_0) =
	\Flimit{F}{z \rightarrow d_0}
	\frac{f(z)-f(d_0)}{S^\alpha_F(z)-S^\alpha_F(d_0)}
	\leq 0.
\]
with the same considerations. The points $c_0$ and $d_0$ can be identified
as points $c$ and $d$ in the statement of the theorem.

If there are no points $y$ such that $f(y)>0$ and neither is the function
zero throughout, then we can choose a point $y$ such that $f(y)<0$
and proceed in a similar manner.
\proofover

\remark
The following example shows that the analogue of Rolle's
theorem can not be made more strict which would have implied existence of a
point $c \in F$ such that $\D^\alpha_F(f(c)) = 0$. Let $C$ be the middle
$\frac{1}{3}$ Cantor set. Define
\begin{eqnarray*}
	f(x) & = S^\alpha_C(x) & 0 \leq x \leq 0.5 \\
	     & = 1 - S^\alpha_C(x) \qquad & 0.5 < x \leq 1.
\end{eqnarray*}
This function satisfies $f(0) = f(1) = 0$. Further,
it is continuous in the interval $[0,1]$. Its set of
change is $C$. The $C^\alpha$-derivative is given by
\begin{eqnarray*}
	\D^\alpha_C(f(x)) & = \chi_C(x) \qquad & 0 \leq x \leq 0.5\\
			& = - \chi_C(x) \qquad & 0.5 < x \leq 1.
\end{eqnarray*}
Thus, $x \in C \Longrightarrow \D^\alpha_C(f(x)) = \pm 1 \neq 0$.

In general, it can be said that the ``fragmented nature'' of the fractal $F$
does not allow us to make the analogue of Rolle's theorem as strict as its
original version.

Now we state the analogue of the law of the mean.
\begin{corollary}
Let $f: \R \rightarrow \R$ be a continuous function such that its set of
change is contained in an $\alpha$-perfect set $F\subset \R$,
$\D^\alpha_F(f(x))$ exists at all points $x\in[a,b]$ and
$S^\alpha_F(b) \neq S^\alpha_F(a)$. Then there exists a point $c \in F$ such
that
\[
	\D^\alpha_F(f(c)) \geq \frac{f(b)-f(a)}{S^\alpha_F(b)-S^\alpha_F(a)}
\]
and a point $d \in F$ such that
\[
	\D^\alpha_F(f(d)) \leq \frac{f(b)-f(a)}{S^\alpha_F(b)-S^\alpha_F(a)}
\]
\end{corollary}
\proof
Let
\[
	g(x) = (f(x) - f(a)) - \frac{f(b)-f(a)}{S^\alpha_F(b)-S^\alpha_F(a)}
	(S^\alpha_F(x)-S^\alpha_F(a))
\]
so that the difference between $f$ and $g$ is a constant plus a multiple of
$S^\alpha_F(x) - S^\alpha_F(a)$. Now $g(a) = g(b) = 0$ so that we can use
the last theorem to
say that there exists a point $c \in F$ such that $\D^\alpha_F(g(c)) \geq 0$
and a point $d \in F$ such that $\D^\alpha_F(g(d)) \leq 0$. This implies that
\[
	\D^\alpha_F(f(c)) - \frac{f(b)-f(a)}{S^\alpha_F(b)-S^\alpha_F(a)}
	\geq 0
\]
for some $c \in F$, and
\[
	\D^\alpha_F(f(d)) - \frac{f(b)-f(a)}{S^\alpha_F(b)-S^\alpha_F(a)}
	\leq 0
\]
for some $d \in F$, which lead to the required relations.
\proofover

We had seen earlier that the $F^\alpha$-derivative of a constant $f(x)=k$ is
zero. Now we see that these are the only functions whose
$F^\alpha$-derivatives are zero:
\begin{corollary}
\label{cor:constant}
Let $f:\R \rightarrow \R$ be a continuous function such that
$\Sch(f) \subset F$ and $\D^\alpha_F(f(x)) = 0$ for all $x \in [a,b]$.
Then $f(x) = k$ where $k$ is a constant on $[a,b]$.
\end{corollary}
\proof
Suppose, if possible, that the function is not a constant. Then there
exist $y$ and $z$,
$y < z$, such that $f(y) \neq f(z)$. This implies either
$f(y) < f(z)$ or $f(y) > f(z)$.

\emph{Case 1}. $f(y) < f(z)$. Then there exists a $c \in F\cap(y,z)$ such
that
\[
	\D^\alpha_F(f(c)) \geq
	\frac{f(z)-f(y)}{S^\alpha_F(z)-S^\alpha_F(y)} > 0.
\]

\emph{Case 2.} $f(y) > f(z)$. Then there exists a $d \in F\cap(y,z)$ such
that
\[
	\D^\alpha_F(f(d)) \leq
	\frac{f(z)-f(y)}{S^\alpha_F(z)-S^\alpha_F(y)} < 0.
\]

In both the cases we have found a point where the derivative is not zero
which contradicts our assumption.~\proofover
\remark
Again due to the ``fragmented nature'' of the fractal $F$, the
$F^\alpha$-differentiability of $f$ is not sufficient to guarantee the
result. Further, The additional conditions that $f$ be a continuous function
and $\Sch(f) \subset F$ are necessary also in the second fundamental
theorem~\ref{thm:funda-two} which relies on the last corollary, and the
integration by parts rule (theorem~\ref{thm:byparts}) which depends on
theorem~\ref{thm:funda-two}.

The $F^\alpha$-derivative satisfies the analogue of Leibniz rule:
\begin{theorem}
\label{thm:leibniz}
If the functions $u: \R \rightarrow \R$ and $v:\R \rightarrow \R$ are
$F^\alpha$-differentiable, then $f(x) = u(x)v(x)$ is
$F^\alpha$-differentiable, and
\be
	\D^\alpha_F(u(x)v(x)) =
	\D^\alpha_F(u(x)) v(x) + u(x)\D^\alpha_F(v(x))
	\label{eqn:leibniz}.
\ee
\end{theorem}
The proof is straightforward.

\section{Fundamental theorems of $F^\alpha$-calculus}
\label{sec:fundamental-theorems}

This section relates the $F^\alpha$-integration and
$F^\alpha$-differentiation as ``inverse processes'' of each other.

The first fundamental theorem says that the $F^\alpha$-derivative is the
inverse of indefinite $F^\alpha$-integral.
\begin{theorem}
\label{thm:funda-one}
Let $F \subset \R$ be an $\alpha$-perfect set.
If $f \in B(F)$ is an $F$-continuous function on $F\cap[a,b]$ and
\[
	g(x) = \int_a^x f(y)\dif^\alpha_F y
\]
for all $x \in [a,b]$, then
\[
	\D^\alpha_F(g(x)) = f(x)\chi_F(x).
\]
\end{theorem}
\proof
If $x \notin F$, then $\D^\alpha_F(g(x)) = 0$ by definition.

For $x \in F$, if there are points in $F$ arbitrarily close to $x$ on both
sides of $x$, then we have to consider both the following cases:

\begin{enumerate}
\item
The set $F \cap (x,z)$ is never empty for $z > x$ and
\[
	g(z) - g(x) = \int_x^{z} f(y)\dif^\alpha_F y.
\]
\item
The set $F \cap (z,x)$ is never empty for $z < x$ and
\[
	g(x) - g(z) = \int_{z}^x f(y)\dif^\alpha_F y.
\]
\end{enumerate}
Otherwise we have to consider only one of the cases which is applicable. We
consider the first one; the second can be treated similarly.

In the first case, $F\cap(x,z)$ is not empty for any $z > x$.
Taking the $F$-limit as $z \rightarrow x$, we get
\be
	\D^\alpha_F(g(x)) = \Flimit{F}{z \rightarrow x}
	\frac{\int_x^{z} f(y)\dif^\alpha_F y}{S^\alpha_F(z)-S^\alpha_F(x)}.
	\label{eqn:funda-A}
\ee
Now,
\[
\fl
	m[f,F,[x,z]] \int_x^{z} \chi_F(y)\dif^\alpha_F y
	\leq \int_x^{z} f(y)\dif^\alpha_F y
	\leq M[f,F,[x,z]] \int_x^{z} \chi_F(y)\dif^\alpha_F y,
\]
and
\[
	\int_x^{z} \chi_F(y)\dif^\alpha_F y = (S^\alpha_F(z)-S^\alpha_F(x))
\]
so that
\be
	m[f,F,[x,z]] \leq
	\frac{\int_x^{z} f(y)\dif^\alpha_F y}{S^\alpha_F(z)-S^\alpha_F(x)}
	\leq M[f,F,[x,z]].
	\label{eqn:funda-B}
\ee
As $f$ is $F$-continuous,
\be
	\Flimit{F}{z \rightarrow x} m[f,F,[x,z]]
	= \Flimit{F}{z \rightarrow x} M[f,F,[x,z]]
	= f(x)
	\label{eqn:funda-C}
\ee
From \eref{eqn:funda-A}, \eref{eqn:funda-B} and~\eref{eqn:funda-C}, we get
the required result.
\proofover

The second fundamental theorem says that the $F^\alpha$-integral as a
function of upper limit is the inverse of $F^\alpha$-derivative except for
an additive constant.
\begin{theorem}
\label{thm:funda-two}
Let $f: \R \rightarrow \R$ be a continuous, $F^\alpha$-differentiable
function such that
$\Sch(f)$ is contained in an $\alpha$-perfect set $F$ and
$h : \R \rightarrow \R$ be $F$-continuous, such that
\[
	h(x)\chi_F(x) = \D^\alpha_F(f(x)).
\]
Then
\[
	\int_a^b h(x)\dif^\alpha_F x = f(b) - f(a).
\]
\end{theorem}
\proof
If
\[
	g(x) = \int_a^x h(x)\dif^\alpha_F x
\]
then $\D^\alpha_F(g(x)) = h(x)\chi_F(x)$ by the last theorem. Therefore
$\D^\alpha_F(g(x)-f(x)) = 0$ for all $x \in [a,b]$. Now
corollary~\ref{cor:constant} implies that
$g(x)-f(x)=k$, a constant, or $g(x) = f(x) + k$. Thus,
\[
	f(b) - f(a) = g(b) - g(a)
	= g(b)
	= \int_a^b h(x)\dif^\alpha_F x
\]
which proves the theorem.
\proofover

The following theorem states that the $F^\alpha$-integration can be
performed by parts, and can be proved by using
fundamental theorem~\ref{thm:funda-two} and Leibniz rule
(theorem~\ref{thm:leibniz}):
\begin{theorem}
\label{thm:byparts}
Let the functions $u:\R \rightarrow\R, v:\R \rightarrow \R$ be such
that
\begin{enumerate}
\item
	$u$ is continuous on $[a,b]$ and $\Sch(u) \subset F$,
\item
	$\D^\alpha_F(u)$ exists and is $F$-continuous on $[a,b]$,
\item
	$v$ is $F$-continuous on $[a,b]$.
\end{enumerate}
Then,
\be
	\fl
	\label{eqn:byparts}
	\int_a^b u(x) v(x)\dif^\alpha_F x = 
	\left[ u(x) \int_a^x v(x')\dif^\alpha_F x'\right]_a^b -
	\int_a^b \D^\alpha_F (u(x)) \int_a^x v(x')\dif^\alpha_F x'
	\dif^\alpha_F x.
\ee
\end{theorem}
The proof is straightforward and omitted.

In \ref{sec:repeated}, we discuss examples of repeated
$F^\alpha$-derivatives and $F^\alpha$-integrals. There we also calculate
$F^\alpha$-derivatives and $F^\alpha$-integrals of powers
$(S^\alpha_F(x))^n$.

Now that the analogies between $F^\alpha$-calculus and ordinary calculus
have become clear, we summarise some of them in \sref{sec:analogies} for a
quick reference.

\section{Examples and applications of $F^\alpha$-Differential equations}
\label{sec:examples}

In this section we briefly touch a couple of examples of
$F^\alpha$-differential equations. The $F^\alpha$-differential equations is
the main topic of a subsequent work~\cite{AP-ADG-sub}.

Firstly we revisit the local fractional diffusion equation proposed
in~\cite{Kolwankar1998} and also discussed partly in~\cite{Kolwankar1999}.
This equation is of the form
\be
	\label{eqn:frac-diff-de}
	\D^\alpha_{F,t}(W(x,t))
	= \frac{\chi_F(t)}{2} \frac{\partial^2}{\partial x^2} W(x,t).
\ee
where the density $W$ is defined as a function of two arguments
$(x,t) \in \R \times \R$
and with a slight change of notation $D^\alpha_{F,t}$ denotes the partial
$F^\alpha$-derivative with respect to time $t$, $\chi_F$ being the
characteristic function of $F$. (This equation may be compared with ordinary
diffusion equation
$\frac{\partial W(x,t)}{\partial t}
= D \frac{\partial^2}{\partial x^2} W(x,t)$.) The Riemann integral like
prescription given in~\cite{Kolwankar1998} had enabled one to construct a
new exact solution. This solution is
\be
	\label{eqn:frac-diff-soln}
	W(x,t) =
	\frac{1}{(2\pi S^\alpha_F(t))^\frac{1}{2}}
	\exp\left(\frac{-x^2}{2S^\alpha_F(t)}\right), \qquad
	W(x,0) = \delta(x).
\ee
This can be recognized as a subdiffusive solution, since $S^\alpha_F$ is
known to be bounded by $k t^\alpha$, $k$ constant, in simple cases including
Cantor sets.

An important observation at this stage is that: equations
like~\eref{eqn:frac-diff-de} are examples of fractal-time evolution
processes.

\subsection*{Motion in a fractally distributed medium}
As a second example, we consider one dimensional motion of a particle
undergoing friction. First we recall the equation of motion in a continuous
(i.~e.\ nonfractal) medium. If the frictional force is proportional to the
velocity, the equation of motion can be written as
\be
	\label{eqn:friction-A}
	\frac{\rmd v}{\rmd t} = -k(x) v
\ee
where $k(x)$, the coefficient of friction, may be dependent on the particle
position $x$.

\Eref{eqn:friction-A} can be reexpressed by considering velocity $v$ as a
function of position $x$. The equation can be written as
\[
	\frac{\rmd v}{\rmd x} \frac{\rmd x}{\rmd t} = -k(x) v.
\]
Identifying $\rmd x/\rmd t = v$ and assuming $v \neq 0$, the equation becomes
\be
	\label{eqn:friction-B}
	\frac{\rmd v}{\rmd x} = -k(x)
\ee
which is readily solved by integrating $k(x)$ if $k(x)$, which models the
frictional medium, is smooth.

If the underlying medium is a fractal, then \eref{eqn:friction-B} is
inadequate to model the motion. Instead we propose the
$F^\alpha$-differential equation of the form
\be
	\label{eqn:friction-C}
	\D^\alpha_F(v(x)) = -k(x)
\ee
for this scenario. Here, the set $F$ is the support of $k(x)$ which describes
the underlying fractal medium, and $\alpha$ is the $\gamma$-dimension of
$F$. (If $F$ is not $\alpha$-perfect, then the set $\Sch(S^\alpha_F)$ can be
chosen instead.) The function $k(x)$ may be called fractional
coefficient of friction due to its physical dimensions.

The solution of \eref{eqn:friction-C} is easily seen to be
\be
	\label{eqn:friction-D}
	v(x) = v_0 - \int_{x_0}^x k(x')\dif^\alpha_Fx'
\ee
where $v_0$ and $x_0$ are the initial velocity and position respectively. In
a simple case where $k(x)$ is uniform on the fractal i.~e.\
$k(x) = \kappa \chi_F(x)$ where $\kappa$ is a constant,
\eref{eqn:friction-D} reduces to
\[
	v(x) = v_0 - \kappa (S^\alpha_F(x) - S^\alpha_F(x_0)).
\]
In the extreme cases we obtain back the classical behaviour: (i)~If $F$ is
empty (frictionless case), then $v(x) = v_0$; (ii)~If $F = \R$ (uniform
medium) then $v(x) = v_0 - \kappa(x-x_0)$.

The time dependence of $x$ is given by
\[
	t(x) = \int_{x_0}^x \frac{1}{v(x')}\dif x'
\]
where $t(x)$ is the time required to reach the position $x$.

\section{Concluding remarks}
In this paper we have developed a calculus on fractal subsets of the real
line. This developement involved the identification of the special role
played by staircase functions associated with fractal sets, which may be
compared with the role of independent variable itself in ordinary calculus.
In particular, $F^\alpha$-integrals and $F^\alpha$-derivatives (of
order~$\alpha$, $0<\alpha\leq1$) are defined using staircase functions
for sets $F$ of dimension $\alpha$.
In contrast with the classical fractional calculus, the notions of
$F^\alpha$-derivatives and $F^\alpha$-integrals are specifically tailored
for fractals of dimension $\alpha$ and thus provide suitable operators on
fractals. Further, they reduce to ordinary derivative and Riemann integral
respectively, when $F = \R$ and $\alpha = 1$.

Much of the developement of the $F^\alpha$-calculus
is carried in analogy with the ordinary calculus.
Several results and techniques of ordinary calculus, including the
Leibniz rule, the fundamental theorems of calculus, the technique of
integration by parts etc.\ have analogues in this calculus.
Specifically we have
adopted Riemann approach for $F^\alpha$-integrations. This approach can
possibly be generalised using Kurzweil-Henstock integration
schemes~\cite{Gordon,bartle}. Work is in progress in this direction.

In the process of the developement of the $F^\alpha$-integrals we have
introduced $\alpha$-mass or mass function associated with a fractal subset
$F$ of the real line. This lead us to introduce the $\gamma$-dimension in
\sref{sec:gamma-dimension}. This dimension is finer than the box-dimension.
Though it is not as fine as the Hausdorff dimension, it is specific to the
developement of calculus here and we expect it to be associated naturally
with algorithms and numerical schemes based on the present calculus.

We have also discussed simple models based on $F^\alpha$-differential
equations.  The solutions of $F^\alpha$-differential equations naturally
involve staircase-like functions. Staircase functions such as the Lebesgue
Cantor staircase function are known to be
bounded by sublinear power laws. Also, they ``change'' or ``evolve'' only on
a fractal set.  Thus, this framework may be useful in modelling many cases
of sublinear behaviour, fractal time evolution, fields due to fractal charge
distributions, etc. The $F^\alpha$-differentiability may be used to classify
singular probability distribution functions.

Continuous-time dynamical systems are associated with ordinary differential
equations, and discrete-time
dynamical systems are associated with maps\slash diffeomorphisms. But as
realised in~\cite{Kolwankar1998}, the
dynamical systems associated with $F^\alpha$-differential equations would be
those evolving on fractal subsets of time-axis.
It would also be of great interest to investigate correspondences between
ordinary differential equations
and $F^\alpha$-differential equations. These are explored in a companion
paper~\cite{AP-ADG-sub}.

There are many obvious directions in which considerations of this paper
should be extended. Some of them are mentioned above. Other important
directions would be extensions to multivariable case, developement of
differential equations and variational principles to mention a few. Work is
in progress in these directions.

\ack
Abhay Parvate is thankful to the Council of Scientific and Industrial
Research~(CSIR), India, for financial assistance. The authors wish to thank
Dr.~H.~Bhate for fruitful discussions.

\appendix

\section{$C^\alpha$-Integrating $f(x) = x\,\chi_C(x)$}
\label{sec:x-chi}

As an example of $F^\alpha$-integration, here we calculate
\be
	\label{eqn:x-one}
	g(y) = \int_0^y x\,\chi_C(x)\dif_C^\alpha x
	= \int_0^y x\dif_C^\alpha x
\ee
where $C$ is the middle~$\frac{1}{3}$ Cantor set, and $\alpha = \ln(2)/\ln(3)$
is its $\gamma$-dimension.
The function $f(x) = x\,\chi_C(x)$ is $C$-continuous on $[0,1]$, hence
it is $C^\alpha$-integrable.

The set $P_n = \{x_i=i/n:0 \leq i \leq n\}$ is a subdivision of
$[0,1]$.  For any component $[x_i,x_{i+1}]$ of $P_n$,
$x_i \leq m[f,F,[x_i,x_{i+1}]]$ and
$x_{i+1} \geq M[f,F,[x_i,x_{i+1}]]$ if $F\cap[x_i,x_{i+1}] \neq\emptyset$.
Therefore,
\[
	\uln{g(1)} =
	\lim_{n\rightarrow\infty} \sum_{i=0}^n \left\{
		\frac{i}{n} \left[
			S^\alpha_C\left(\frac{i+1}{n}\right)
			- S^\alpha_C\left(\frac{i}{n}\right)
		\right]
	\right\} \leq L^\alpha[f,F,P_n]
\]
and
\[
	\oln{g(1)} =
	\lim_{n\rightarrow\infty} \sum_{i=0}^n \left\{
		\frac{i+1}{n} \left[
			S^\alpha_C\left(\frac{i+1}{n}\right)
			- S^\alpha_C\left(\frac{i}{n}\right)
		\right]
	\right\} \geq U^\alpha[f,F,P_n].
\]
Further, it can be seen that
\[
	\lim_{n\rightarrow\infty} [\oln{g(1)} - \uln{g(1)}] = 0.
\]
Thus, $g(1)$ can be calculated using the limit
\be
	\label{eqn:x-A-one}
	g(1) =
	\lim_{n\rightarrow\infty} \sum_{i=0}^n \left\{
		\frac{i}{n} \left[
			S^\alpha_C\left(\frac{i+1}{n}\right)
			- S^\alpha_C\left(\frac{i}{n}\right)
		\right]
	\right\}.
\ee
Similarly for integers $m>0$,
\be
	\label{eqn:x-A-two}
	g\left(\frac{1}{3^m}\right) =
	\lim_{n\rightarrow\infty} \sum_{i=0}^n \left\{
		\frac{i}{3^m n} \left[
			S^\alpha_C\left(\frac{i+1}{3^m n}\right)
			- S^\alpha_C\left(\frac{i}{3^m n}\right)
		\right]
	\right\}.
\ee
Using the self-similarity of $C$ and scaling of $S^\alpha_C$, we see from
\eref{eqn:x-A-one} and~\eref{eqn:x-A-two} that
\be
	\label{eqn:x-A-three}
	g\left(\frac{1}{3^m}\right) = \frac{1}{3^{m(1+\alpha)}}\, g(1)
	= \frac{1}{6^m}\, g(1).
\ee

We make use of the ternary representation of numbers which simplifies many
calculations involving the Cantor set. Any number $y \in [0,1]$ can be
represented by the series
\be
	\label{eqn:x-two}
	y = \sum_{i=1}^\infty \frac{t_i(y)}{3^i}
\ee
where $t_i(y) = 0,1\mbox{ or }2$ is the $i$th ternary digit of $y$ after
ternary point. The number $y$ belongs to $C$ if and only if $y$ has a
representation of the form~\eref{eqn:x-two} where $t_i(y) = 0\mbox{ or }2$ for
all~$i$.

An approximation of $y \in [0,1]$ by a finite number of digits is denoted by
\be
	\label{eqn:x-three}
	T_0(y) = 0 \quad\mbox{and}\quad
	T_n(y) = \sum_{i=1}^n \frac{t_i(y)}{3^i}.
\ee
The sequence $\{T_n(y)\}_{n=0}^\infty$ is a monotonically (but not strictly)
increasing sequence whose limit is $y$. Hence we can write
\be
	\label{eqn:x-four}
	g(y) = \sum_{i=1}^\infty \int_{T_{i-1}(y)}^{T_i(y)}
		x\,\chi(x)\dif^\alpha_C x
		= \sum_{i=1}^\infty I_i(y)
\ee
where
\be
	\label{eqn:x-five}
	I_i(y) = \int_{T_{i-1}(y)}^{T_i(y)} x\,\chi(x)\dif^\alpha_C x.
\ee
The quantities $I_i(y)$ can be calculated using the self-similarity of $C$,
the scaling and translation properties of $S^\alpha_C$
(theorem~\ref{thm:scaling}), and
\eref{eqn:x-A-three}. Let $y \in [0,1]$ and let $n$ be any integer such
that $i < n \Longrightarrow t_i(y) = 0\mbox{ or }2$. Then
$i < n \Longrightarrow T_i(y) \in C$. For calculating $I_n(y)$, there are
three cases corresponding to three possible values of $t_n(y)$:
\begin{description}
\item[Case $t_n(y) = 0$:]
	Here, $T_{n-1}(y) = T_n(y)$ and $I_n(y) = 0$.
\item[Case $t_n(y) = 1$:]
	In this case,
	\[
		T_n(y) - T_{n-1}(y) = \frac{1}{3^n}
		=\sum_{i=n+1}^\infty \frac{2}{3^i}
	\]
	so that there is another sequence $\{t_i(T_n(y))\}$ which does not
	contain the digit 1, hence $T_n(y) \in C$. The set
	$[T_{n-1}(y),T_n(y)] \cap C$ can be written as
	\[
		\{z: i<n \Longrightarrow t_i(z) = t_i(y);\ t_n(z) = 0;\
		i>n \Longrightarrow t_i(z) = 0\mbox{ or }2\}
	\]
	Therefore it is a scaled down version of $C$ by a factor $1/3^n$ and
	translated by $T_{n-1}(y)$. Hence writing
	$x = T_{n-1}(y) + (x - T_{n-1}(y))$, we get
	\begin{eqnarray*}
	\fl I_n(y)
	= T_{n-1}(y) \int_{T_{n-1}(y)}^{T_n(y)} \chi_C(x)\dif^\alpha_C x
		+ \int_{T_{n-1}(y)}^{T_n(y)}
		(x - T_{n-1}(y))\,\chi_C(x)\dif^\alpha_C x \\
	\lo= T_{n-1}(y) \int_0^{1/3^n} \chi_C(x)\,d^\alpha_C x
		+ \int_0^{1/3^n} x\,\chi_C(x)\dif^\alpha_C x \\
	\lo= T_{n-1}(y) \frac{1}{\Gamma}\frac{1}{3^{\alpha n}}
		+ \frac{1}{3^{n(1+\alpha)}}\,g(1) \\
	\lo= \frac{T_{n-1}(y)}{\Gamma 2^n} + \frac{g(1)}{6^n}
	\end{eqnarray*}
	where $\Gamma$ denotes $\Gamma(\alpha+1)$ for convenience.
	If $y = T_n(y)$ then $y \in C$. But if $y > T_n(y)$, then as
	$t_n(y)=1$ and $t_i\neq 0$ for some $i > n$, therefore $y \notin C$.
	Thus the half open interval $(T_n(y),y]$ does not intersect $C$
	implying that $I_k(y) = 0$ for all $k > n$.
\item[Case $t_n(y) = 2$:]
	Here, $T_n(y)$ clearly belongs to $C$. If $D$ is the set
	\[ \fl
		D = \{z: i<n \Longrightarrow t_i(z) = t_i(y);\
		t_n(z) = 0;\ i>n \Longrightarrow t_i(z) = 0\mbox{ or }2\}
	\]
	then $D$  is a scaled down version of $C$ by a factor $1/3^n$,
	$D\subset[T_{n-1}(y),T_n(y)]$, and more specifically,
	$[T_{n-1}(y),T_n(y)]\cap C = D \cup \{T_n(y)\}$. Therefore by
	arguments similar to the case $t_n(y) = 1$,
	\be
		\label{eqn:x-six}
		I_n(y) = \frac{T_{n-1}(y)}{\Gamma 2^n} +
		\frac{g(1)}{6^n}
	\ee
	But unlike the case $t_n(y) = 1$, there is a possibility that
	$C\cap(T_n(y),y]$ is nonempty so that $I_k(y)$ need not be zero for
	all $k > n$.
\end{description}
Summarizing,
\be
	\label{eqn:x-seven}
	I_n(y) = \cases{
		0 & if $t_n(y) = 0$ or\\
		& $t_i(y) = 1$ for some $i < n$\\
		\frac{T_{n-1}(y)}{\Gamma 2^n}
		+ \frac{g(1)}{6^n}
		& otherwise.
	}
\ee
This description requires the value of $g(1)$. It can be found out by
putting $y = 1$ in \eref{eqn:x-seven}. If $y = 1$, then $t_i(y)=2$ for
all~$i$. Also,
\[
	T_n(1) = \sum_{i=1}^n \frac{2}{3^i} = 1 - 3^{-n}.
\]
Therefore,
\[
	I_n(1) = \frac{1-3^{-(n-1)}}{\Gamma 2^n}
	+ \frac{g(1)}{6^n}.
\]
Substituting this in \eref{eqn:x-seven} and solving \eref{eqn:x-four}
for $g(1)$, we get
\be
	\label{eqn:x-eight}
	g(1) = \frac{1}{2\Gamma}.
\ee
Thus,
\be
	g(y) = \int_0^y x \chi_C(x)\dif^\alpha_F x
	= \sum_{n=1}^\infty I_n(y)
	\label{eqn:x-pre-nine}
\ee
where
\be
	\label{eqn:x-nine}
	\fl I_n(y) = \cases{
		0 & if $t_n(y) = 0$ or\\
		& $t_i(y) = 1$ for some $i<n$\\
		\frac{1}{\Gamma(\alpha+1)}
		\left[\frac{T_{n-1}(y)}{2^n} +
		\frac{1}{2\cdot6^n}\right]
		& otherwise
	}
\ee
and $T_n(y)$ are given by equations~\eref{eqn:x-two} and~\eref{eqn:x-three}.

\section{Regarding repeated $F^\alpha$-integration and
$F^\alpha$-derivative}
\label{sec:repeated}

\subsection*{$F^\alpha$-derivative}
Many dynamical systems are modelled by differential equations involving
second and higher order
derivatives i.~e.\ derivative operator applied repeatedly. The successive
operation of the $\D^\alpha_F$ operator is also possible and gives
meaningful results. As an example, let us $F^\alpha$-differentiate the
function $g(x) = (S^\alpha_F(x))^2$ twice, where
$F \subset R$ is an $\alpha$-perfect set. By definition of the derivative,
\be
	x \notin F \Longrightarrow \D^\alpha_F g(x) = 0.
	\label{eqn:sq-A}
\ee
If $x \in F$, then
\begin{eqnarray}
\D^\alpha_F g(x)
& = \Flimit{F}{y \rightarrow x}
	\frac{(S^\alpha_F(x))^2 - (S^\alpha_F(x))^2}%
	{S^\alpha_F(y) - S^\alpha_F(x)} \nonumber \\
& = 2\, S^\alpha_F(x). \label{eqn:sq-B}
\end{eqnarray}

Equations~(\ref{eqn:sq-A}) and~(\ref{eqn:sq-B}) can be combined to give
\be
	\D^\alpha_F\ (S^\alpha_F(x))^2 = 2\,S^\alpha_F(x)\,\chi_F(x)
	\label{eqn:sq-C}
\ee
where $\chi_F$ is the characteristic function of $F$. As a side remark, it
is easy to generalize this to
\be
	\label{eqn:s-n-d}
	\D^\alpha_F((S^\alpha_F(x))^n) = n (S^\alpha_F(x))^{n-1}\chi_F(x)
	\label{eqn:sq-Cb}
\ee
for any integer $n>0$.

Now we take the second $F^\alpha$-derivative of $g$. As far as the
operator~$\D^\alpha_F$ is concerned, the values of the function outside $F$
make no difference because of the $F$-limit in its definition. Thus,
\begin{eqnarray}
(\D^\alpha_F)^2 (S^\alpha_F(x))^2
& = \D^\alpha_F (2\,S^\alpha_F(x)\,\chi_F(x)) \nonumber \\
& = 2\,\D^\alpha_F\,S^\alpha_F(x) \nonumber \\
& = 2\,\chi_F(x) \label{eqn:sq-Ca}
\end{eqnarray}
Where the last step follows from lemma~\ref{lem:staircase-derivative} and
linearity (theorem~\ref{thm:derivative-linearity}).

Apart from the $\gamma$-dimension of $F$, the order $\alpha$ also has
another significance. This will be clear from the following example. If $C$
is the Cantor set, then it is known~\cite{hille} that $S^\alpha_C(x)$
is bounded by the power $\alpha$ of $x$ from below and above:
\be
	a x^\alpha \leq S^\alpha_C(x) \leq b x^\alpha
\ee
where $\alpha = \ln(2)/\ln(3)$ is the $\gamma$-dimension of $C$. Hence the
function $g$ defined above is bounded by power $2\alpha$ of $x$:
\be
	a x^{2\alpha} \leq g(x) = (S^\alpha_C(x))^2 \leq b x^{2\alpha}
\ee
so that
\be
	x \in F \Longrightarrow
	2a x^\alpha \leq \D^\alpha_C (g(x)) \leq 2b x^\alpha
\ee
and
\be
	x \in F \Longrightarrow
	(\D^\alpha_C)^2 (g(x)) = 2.
\ee
This example demonstrates that $F^\alpha$-differentiation reduces the
power of bounds by $\alpha$.

\subsection*{$F^\alpha$-integration}
The $F^\alpha$-integration can also be carried out in succession. Let $F$ be
an $\alpha$-perfect set.  It is already shown that $F^\alpha$-integration of
$\chi_F(x)$ is $S^\alpha_F(x)$:
\be
	\int_a^{x'} \chi_F(x)\dif^\alpha_F x
	= S^\alpha_F(x')
\ee
where for simplicity we have taken $S^\alpha_F(a) = 0$ which is consistent
with the definition~\ref{def:staircase} of the staircase function. Now,
\[
	g_1(x') \equiv \int_a^{x'} S^\alpha_F(x)\dif^\alpha_F x
	= \frac{1}{2} (S^\alpha_F(x'))^2
\]
where we have used \eref{eqn:sq-C} and the fundamental
theorem~\ref{thm:funda-two}. Again, it is easy to generalise this to
\be
	\label{eqn:s-n-i}
	\int_a^{x'} (S^\alpha_F(x))^n \dif^\alpha_F x
	= \frac{1}{n+1} (S^\alpha_F(x'))^{n+1}
\ee
using \eref{eqn:s-n-d}.

\section{A few analogies between $F^\alpha$-calculus and ordinary calculus}
\label{sec:analogies}

The $F^\alpha$-calculus can be thought of as a generalization of ordinary
calculus with Riemann approach. \Tref{tbl:analogies} shows a few
analogies between various quantities.
\Table{\label{tbl:analogies}A few analogies between $F^\alpha$-calculus
and ordinary calculus.}
\br
Ordinary calculus & $F^\alpha$-calculus \\
\mr
$\R$ & An $\alpha$-perfect set $F$ \\
limit & $F$-limit \\
Continuity & $F$-continuity \\
$\ds \int_0^y x^n\dif x = \frac{1}{n+1} y^{n+1}$
	& $\ds \int_0^y (S^\alpha_F(x))^n\dif^\alpha_F x
	= \frac{1}{n+1} (S^\alpha_F(y))^{n+1}$\\
$\ds \frac{\rmd}{\rmd x} x^n = n\,x^{n-1}$ &
	$\ds \D^\alpha_F((S^\alpha_F(x))^n)
	= n\,(S^\alpha_F(x))^{n-1}\,\chi_F(x)$\\
Leibniz rule & Theorem~\ref{thm:leibniz} \\
Fundamental theorems &
	Theorems~\ref{thm:funda-one} and~\ref{thm:funda-two}\\
Integration by parts & Theorem~\ref{thm:byparts}\\
\br
\endTable


\end{document}